\documentclass{elsart}
\usepackage{ifpdf}
\usepackage{graphicx,natbib,amssymb,lineno,subfigure}
\ifpdf
\usepackage[%
  pdftitle={},%
  pdfauthor={},%
  pdfsubject={},%
  pdfkeywords={},%
  pdfstartview=FitH,%
  bookmarks=true,%
  bookmarksopen=true,%
  breaklinks=true,%
  colorlinks=true,%
  linkcolor=blue,anchorcolor=blue,%
  citecolor=blue,filecolor=blue,%
  menucolor=blue,pagecolor=blue,%
  urlcolor=blue]{hyperref}
\else
\usepackage[%
  breaklinks=true,%
  colorlinks=true,%
  linkcolor=blue,anchorcolor=blue,%
  citecolor=blue,filecolor=blue,%
  menucolor=blue,pagecolor=blue,%
  urlcolor=blue]{hyperref}
\fi

\makeatletter
\def\elsartstyle{%
    \def\normalsize{\@setfontsize\normalsize\@xiipt{14.5}}
    \def\small{\@setfontsize\small\@xipt{13.6}}
    \let\footnotesize=\small
    \def\large{\@setfontsize\large\@xivpt{18}}
    \def\Large{\@setfontsize\Large\@xviipt{22}}
    \skip\@mpfootins = 18\p@ \@plus 2\p@
    \normalsize
}
\@ifundefined{square}{}{}
\makeatother

\renewcommand{\vec}[1]{{\bf #1}}

\begin{document}

\begin{frontmatter}
  \title{The classical granular temperature\\ and slightly beyond}

\author[TAU]{D. Serero},
\ead{serero@eng.tau.ac.il}
\author[PMMH]{C. Goldenberg},
\ead{chayg@pmmh.espci.fr}
\thanks{I.~G. gratefully acknowledges support from the Israel Science Foundation
(ISF), grant. no. 689/04, the German Israel Science Foundation
(GIF), grant no. 795/2003 and the US-Israel Binational Science Foundation
(BSF), grant no. 2004391. CG gratefully acknowledge support from a European Community FP6 Marie Curie Action
  (MEIF-CT2006-024970).}
\author[TAU]{S. H. Noskowicz},
\ead{henri@eng.tau.ac.il}
\author[TAU]{I. Goldhirsch}
\ead{isaac@eng.tau.ac.il}

\address[TAU]{Department of Fluid Mechanics and Heat Transfer,
Faculty of Engineering,\\ Tel Aviv University,
Ramat-Aviv, Tel Aviv 69978, Israel.}
\address[PMMH]{Laboratoire de Physique et M\'ecanique des Milieux
  H\'et\'erog\`enes\\ (CNRS UMR 7636), ESPCI, 10 rue Vauquelin, 75231 Paris Cedex
  05, France.}

\begin{abstract}
One  goal of this paper is to discuss the classical  definition of granular
temperature as an extension of its  thermodynamic equivalent and  a
 useful
concept   which provides an important characterization of fluidized 
granular matter.  Following a review of some basic
concepts and techniques (with emphasis on fundamental issues) we present
new results for a system that can exhibit strong violations of equipartition,
yet  is amenable to description by classical granular hydrodynamics,
namely a binary granular gas mixture.  A second goal of this article is
to present a  result that pertains to dense granular and molecular solids alike,
namely the existence of 
 a correction to the elastic
energy which is related to the heat flux in the equations of continuum
mechanics.   The latter  is of the same (second) order in the strain as the
elastic energy.  Although  recent definitions of temperatures for 
granular matter, glasses and other disordered many-body systems are not 
within the scope of this article we do make several general comments on this
subject in the closing section. 
\end{abstract}
\end{frontmatter}
\section{Introduction}
\label{intro1}

The axiomatic formulation of thermodynamics provides a precise definition of
 temperature. Nearly any textbook on thermodynamics shows this definition
to be compatible with common experience. 
Equilibrium statistical mechanics  endows temperature with a  microscopic
meaning but invokes  a few assumptions (e.g., ergodicity). 
The physical  relevance and usefulness 
 of both of these theories is beyond debate.

 As there are practically no equilibrium systems in nature there is clearly a
 need to describe non-equilibrium systems.  The most successful theories that
 treat such systems are those devoted to near-equilibrium states, namely states
 in which the macroscopic fields vary sufficiently slowly in space and time so
 that (near) local equilibrium distributions can establish themselves
 everywhere while the state of the system hardly changes. That this should
 occur is not entirely obvious. It helps to note that e.g., in gases local
 equilibration occurs within a few collisions per particle, i.e. on time scales
 that are much smaller than macroscopic times.  In other words, these
 approaches are effective when there is significant scale separation between
 the spatial and temporal micro-and-macroscales.  In the latter case one can
 justify hydrodynamics, elasticity and other macroscopic theories, and the
 transport coefficients are given by Green-Kubo expressions derived using
 linear (and non-linear) response theories \cite{IO} or projection operator
 approaches \cite{Mori1,Mori2}, a similar statement holding for fluctuation
 dissipation `theorems' e.g., for Brownian motion \cite{vanKampen86}.

 The macroscopic fields that represent the states of systems that possess scale
 separation are usually taken to be the densities of the conserved entities,
 namely the (mass or number) density, the momentum density and the energy
 density. These fields are ``slow'' since any change in their local values
 necessitates motion to (non-locally) transport an amount of the entity
 represented by the field.  It is usually assumed that on macroscopic time
 scales the system `remembers' only the slow fields, the other degrees of
 freedom rapidly accommodating themselves to those fields and becoming
 ``noise'' whose statistics is determined by the slow fields.  The description
 of far-from-equilibrium systems, such as turbulent fluids and glasses, is not
 as intuitively founded as that of near-equilibrium systems; in particular,
 since many of these systems lack scale separation it is a-priori unclear which
 fields are sufficient to provide a closed description for them (nor is the
 nature of a relevant statistical ensemble for the unresolved degrees of
 freedom obvious).

The BBGKY hierarchy (see e.g., \cite{HarrisX}) is an exact formulation for
many-body systems that does not require them to be reversible
\cite{Goldhirsch07a} on the
microscopic (particle scale) level or characterized by `slow' or 
hydrodynamic fields.  However, without approximations that
truncate it to a manageable size it is nearly useless. A truncation
which is valid at low densities (gases) begets the  Boltzmann equation
\cite{HarrisX,Chapman70}. Like the BBGKY hierarchy, the Boltzmann equation
is well defined without the introduction of the concept of temperature
or other macroscopic fields and it resolves near-microscopic  scales. 
However it is a nontrivial integrodifferential
equation that can be directly solved in very few cases. 
 One of the important uses of the Boltzmann equation is the 
derivation of hydrodynamic constitutive relations. The form of the
hydrodynamic equations (i.e., the equations of continuum mechanics) is 
very general and holds  for single realizations (see Appendix A).  They also follow from
the Boltzmann equation (whose unknown is a distribution function).
A gradient (Chapman-Enskog) expansion of the 
solution of the Boltzmann equation  begets
explicit expressions for the hydrodynamic fluxes in terms of 
gradients of the hydrodynamic fields (constitutive relations). 
This  expansion is formally valid when there is scale separation
(see more below) although the resulting hydrodynamic equations have been
successfully employed  when scale separation is weak or non-existent, such as
in the case of weak shocks or granular matter \cite{Goldhirsch03}.

Granular systems are different from molecular
systems not only in the respective sizes of their  constituents
 but (more importantly)  the fact that the grains  interact in a dissipative way
(e.g., collide inelastically in the granular gas). Therefore all granular
states are of non-equilibrium nature. An important consequence of this fact \cite{Tan98} is that granular systems do not possess strong scale
separation except in the near-elastic
case. In addition, typical granular systems comprise much fewer than
(say) an  Avogadro
number of grains, therefore they  are very small by molecular standards.
Therefore granular systems are basically 
{\it mesoscopic} \cite{Tan98}: they  are expected to exhibit strong
fluctuations within any reasonably defined ensemble and 
 realizations  may not be  representative of these ensembles, see e.g.,
\cite{Goldenberg02,Goldenberg06b}, even in the fluidized phase.

Imagine a granular gas which so close to being elastic that the percentage of
the energy lost in a collision is minute. Since local equilibration is
established in a matter of a few collisions per particle the main effect of the
weak inelasticity is to cause the decay of the kinetic energy on long time
scales (of course, if the system is forced this `lost' energy can be
replenished). Therefore near-elastic granular gases are expected to behave in a
way that is close to that of elastic gases, as detailed calculations show
\cite{Goldhirsch03}.  In other words, the elastic limit is not singular (for
finite times) and one expects the temperature to be a relevant field with
properties that are similar to those encountered in elastic systems. It is
a-priori unclear whether a similar statement holds for strongly inelastic
granular gases. It is interesting to note that even in the realm of
near-elastic granular gases there is no known equivalent of Boltzmann's
H-theorem.

Granular solids are jammed in static positions with no possibility to explore
phase space. They are athermal. Furthermore, their states are typically
metastable (the ground state of a sand-pile is one in which all grains reside
on the ground).  Although the concept of temperature seems to be a-priori
irrelevant to these systems, there have been interesting suggestions to define
effective (configurational) temperatures for them as properties of appropriate
ensembles, e.g., \cite{Mehta89,Edwards90}. Slow or quasistatic flows of dense
granular matter are clearly athermal as well. Here possible dynamic
applications of the above mentioned effective temperatures, such as
fluctuation-dissipation (FD) relations have been proposed, and tested, see
e.g., \cite{Makse02,Mayor05}, indicating a possible dynamical relevance of
these temperatures. Unlike the FD relations for granular gases
\cite{Goldhirsch00,Brilliantov05,Dufty06} (which are not trivial but can be
derived by means borrowed from near-equilibrium statistical mechanics), the FD
relations for dense and quasistatic granular flows (or glasses) have been
conjectured (see, though \cite{Puglisi02}). Still the above mentioned as well
as other findings are interesting and suggestive of a possible underlying
theoretical framework. Such a framework is probably not going to enjoy the
generality of the near-equilibrium formulations and may necessitate an increase
in the number of variables that characterize the considered systems (as is
often the case in non-equilibrium situations).  A proposal for defining
non-equilibrium temperatures which has attracted considerable attention is due
to Tsallis \cite{Tsallis88}. Whether this definition is relevant to granular
matter is unclear as yet, see though \cite{Beck06}.

In  old theories of granular matter (gases included) the
granular  temperature was not
accounted for (as it  does not correspond to a conserved entity) \cite{Hutter94}.  Even in 
some modern studies  the granular temperature is (used but)
 eliminated from
the equations of motion {[}e.g.,  for  length
scales for which heat conduction is subdominant to heat production
and dissipation \cite{Kumaran06}{]}; in these cases  the temperature is 
claimed to be enslaved to the other fields (but it still  plays a useful
role as a state characterizing field).
Systematic derivations of hydrodynamic equations for granular gases
\cite{Sela98,Brey98} invariably account for the granular temperature as a
relevant field. These equations have been validated 
against simulations and experiments \cite{Goldhirsch03}, even when scale
separation is weak.

The present paper has three major goals. The first is to show that the
temperature field is needed for the description of granular gases as a
characterizing field,  if not a quasi-thermodynamic entity. This
is done in Section 2, where the concept of 
granular temperature is explained in  the context
of the CE expansion. 
 The goal of  Section 3  is to provide an example of 
a kinetic theory  (for binary granular gases)  
 in which the temperature field is
one of the hydrodynamic fields and which does  produce sensible results even when 
equipartition is strongly violated (here a novel  
 method for carrying out the 
CE expansion is presented and applied).  The
third goal, taken up in Section 4, 
 is to show the existence of a correction to the classical formula
for the elastic energy of granular and molecular matter alike; this
correction is shown to be related to the heat flux in the equations of 
continuum mechanics. 
 Section 5 concludes this paper and briefly discusses some issues concerning
non-thermodynamic temperatures. Some technical details
are relegated to Appendices.

\section{Granular Gases and the Chapman-Enskog expansion}
\label{section2}
As mentioned in the Introduction,  the  equations of continuum mechanics 
 are very general and 
in particular, they  hold for practically all single 
realizations of many body systems (see Appendix A). These equations comprise the equation of continuity  (for the density, ${\bf \rho}$), Eq.~(\ref{rhodot1}), 
the momentum density, ${\bf p}$,  equation,  Eq.~(\ref{velvel}), 
and the equation for the energy density, $e$, 
 Eq.~(\ref{energy}), in which 
 (in the dissipative/inelastic case) 
an energy sink term,  ${\bf \rho \Gamma}$, arises. An equation for the 
internal specific energy easily follows from the above, 
see Eq.~(\ref{internal}). Note that the  derivation in Appendix A
produces also explicit
expressions for the relevant  fields and fluxes in terms of the microscopic
degrees of freedom. In particular, note that the expression for the stress
tensor is not precisely the same as the commonly used Voigt or Born or Kirkwood
expression; the latter are limits of Eq.~(\ref{stress}) when the coarse-graining
volume (or width of the coarse-graining function) is very large and do not
satisfy the equations of continuum mechanics unless there is scale separation
or one considers homogeneous deformations of lattice configurations. 

In the realm of molecular or granular gases, when the internal potential 
energy is negligible (for dilute systems) or when a model of hard spheres
is employed one can replace the specific internal energy by
the (granular) temperature,  $ T$.
In this case  Eq.~(\ref{internal}) can  be rewritten as:
\begin{equation}
\rho({\bf r},t) \frac{DT({\bf r},t)}{Dt}
= \frac{\partial V_\beta({\bf r},t)}{\partial r_\alpha} \sigma_{\beta\alpha}
({\bf r},t) - \mathrm{div}\, {\bf Q}({\bf r},t) - \rho({\bf r},t)\Gamma({\bf r},t) \label{Temp}
\end{equation}
with notation that is defined in Appendix A. 
Following Eq.~(\ref{defT})  the (granular)
temperature field is given by the following intuitive expression: 
\begin{equation}
T({\bf r},t)= \frac{\frac{1}{2}\sum_i m_i v^{\prime 2}_i(t) \phi\left[ {\bf r} - {\bf r}_i(t)  \right]}{\sum_j m_j \phi\left[ {\bf r} - {\bf r}_j(t)\right]}
\end{equation}
Note that when scale separation does not exist,  the
temperature, much like other fields,  may be resolution dependent 
(i.e. it
depends on the width of the coarse graining function $\phi$)\cite{Glasser01}.

%\subsection{On granular hydrodynamics}

The  hydrodynamic  description of a fluid, granular or otherwise,
is complete when constitutive relations express  $\mbox{\boldmath$\sigma$}$, ${\bf Q}$ and
$\Gamma$ in terms of the hydrodynamic fields. 
It is {\it a-priori}
unclear whether such a closure exists for granular fluids. Extended
Chapman-Enskog (CE)  expansions provide explicit 
constitutive relations for granular gases, as further detailed below.

The Boltzmann equation is a 
non-trivial integrodifferential equation for the single-particle velocity
distribution, $f({\bf v},{\bf r},t)$, whose integral over the (particle)
velocities, ${\bf v}$,  is the number density, $n({\bf r},t)$. The ratio 
$f({\bf v},{\bf r},t)/n({\bf r},t)$ is the probability density for a particle
to have velocity ${\bf v}$ when it is found (or more accurately its center
of mass is found) at point ${\bf r}$ at time $t$. In the relatively simple 
case of a dilute monodisperse collection of smooth hard 
spheres of unit mass and diameter
$d$ whose collisions are characterized by a fixed coefficient of normal
restitution, $\alpha$, the Boltzmann equation reads
\cite{Sela98,Brey98,Goldshtein95}:
\begin{displaymath}
\frac{\partial f({\bf v}_1)}{\partial t}+{\bf v_{1}}\cdot\mbox{\boldmath$\nabla$}f ({\bf v}_1)
\end{displaymath}
\begin{equation}
=
d^2 \int_{{\bf \hat{k}}\cdot{\bf v_{12}}>0}d{\bf v_{2}}
d{\bf \hat{k}}({\bf \hat{k}}\cdot{\bf v_{12}})
\left(\frac{1}{{\alpha}^2}f({\bf v'_{1}})f({\bf v'_{2}})
-f({\bf v_{1}})f({\bf v_{2}})\right)\label{Boltzmann101}
\end{equation}
where  $\mbox{\boldmath$\nabla$}$ is a spatial gradient. The unit vector $\hat{k}$ points from the center of 
particle `1' to the center of particle `2'. The relation between
the  precollisional (primed) and postcollisional 
velocities  is given by:
${\bf v_{i}}={\bf v'_{i}}-\frac{1+\alpha }{2}({\bf \hat{k}}\cdot{\bf v'_{ij}})
{\bf \hat{k}}$
where  ${\bf v'_{ij}}\equiv{\bf v'_{i}}-{\bf v'_{j}}$. 
The dependence of $f$
on the spatial  coordinates  and time is not explicitly spelled out in 
Eq. (\ref{Boltzmann101}).
Notice that in addition to the explicit dependence of Eq. (\ref{Boltzmann101}) on $\alpha$, it also
implicitly depends on $\alpha$ through the 
relation between the postcollisional and precollisional velocities. The
condition $\hat{k}\cdot {\bf v}_{12}> 0$ represents the fact that only particles whose relative velocity is such that they approach each other, can collide. 

 In the
derivation of the Boltzmann equation \cite{HarrisX,Chapman70} it is assumed
 that  the velocities and positions of 
colliding particles are not correlated. This assumption of ``molecular
chaos'' is
justified for dilute  gases but not  for dense gases or 
 strongly inelastic granular gases \cite{Tan98,Goldhirsch93a,Soto01,Luding00}. This fact is related to the reduction in the normal component
of the relative velocity of colliding particles. In relatively dense gases
 particles heading towards
each other may not collide as they are `stopped' by collisions with
other particles. This phenomenon is modeled  in the
Enskog-corrected Boltzmann equation \cite{Chapman70}. 

The Chapman-Enskog expansion is a method to perturbatively solve the 
Boltzmann equation. Scale separation is formally 
necessary for its implementation.  
When all gradients vanish, a local Maxwellian distribution, corresponding to the values of the hydrodynamic fields at each point, solves
the Boltzmann equation in the elastic case. A gradient expansion (formally, an expansion in the
Knudsen number) around this solution automatically yields the result
that the 
distribution function depends on space and time only through its dependence
on the fields (which is  quasi-local since $f({\bf r},t)$  depends
on the values of the fields and their spatial derivatives at 
$({\bf r},t)$). The often made statement that within the CE expansion
 $f$ is assumed to depend on space and time  only through its
dependence on the hydrodynamic fields is not precise: this assumption is made only for the zeroth order of the expansion.

In the case of granular gases, the 
(kinetic) energy density  is not strictly conserved, and this  led
to questions whether it should be included in the set of hydrodynamic
fields  \cite{Goldhirsch03,Kumaran06,Noije00,Wakou02}.
Following the  discussion in the introduction, when  the degree
of inelasticity is sufficiently small, it is justified to include  the 
granular temperature in the set of 
hydrodynamic fields. An extension of the Chapman Enskog expansion
based on this approach has been proposed and implemented in \cite{Sela98}.
There, the small parameters used to expand the Boltzmann equation were
the Knudsen number,
 $K\equiv\frac{\ell}{L}$,  where $\ell$ is the mean free
path given by $\ell=\frac{1}{\pi n d^2}$,  and $L$ is a macroscopic length
scale (i.e.,  the length scale which is resolved by hydrodynamics, not
necessarily the system size) {\it and} the degree of inelasticity, defined
by  $\epsilon\equiv 1-{\alpha}^2$. In another approach \cite{Brey98} to the problem
only the Knudsen number is the small parameter and the temperature is assumed
to comprise a relevant hydrodynamic field for all values of the 
degree of inelasticity. This method does not seem to lead to contradictions
for any value of $\alpha$; as a matter of fact it
has recently been used to obtain accurate values for the linear transport
coefficients for all physically allowed values of $\alpha$ 
\cite{Noskowicz06}. Whether the results are  indeed
relevant to  strongly inelastic granular gases, where scale separation
does not hold and precollisional correlations may be prominent,  remains to
be seen.

\section{Binary granular gases: extreme violations of equipartition}

Consider  a mixture of smooth hard spheres, composed of species  $A$ 
and  $B$, of masses   $m_{A}$ and $m_{B}$, and diameters
$\sigma_{A}$ and $\sigma_{B}$, respectively. 
The coefficient of normal restitution (assumed to be fixed) for a collision of a particle of
species $\alpha \in \{ A,B \}$ with a particle of species $\beta \in 
\{ A,B\}$ is denoted by $e_{\alpha\beta}$ (hence, $e_{\alpha\beta}
\in \{ e_{AA}, \ e_{BB}, \ e_{AB} \} $).
The transformation of velocities due to a collision of a sphere of species
$\alpha$ with a sphere of species $\beta$ is  given by:
\begin{eqnarray}
\mathbf{v}_{1} &=&\mathbf{v}_{1}^{\prime }-\left( 1+e_{\alpha \beta }\right)
M^{\alpha\beta}\left( \mathbf{v}_{12}^{\prime }\cdot \hat{k}\right) \hat{k}
\label{vtransfo} \\
\mathbf{v}_{2} &=&\mathbf{v}_{2}^{\prime }+\left( 1+e_{\alpha \beta }\right)
M^{\alpha\beta}\left( \mathbf{v}_{12}^{\prime }\cdot \hat{k}\right) \hat{k}  
\end{eqnarray}
where $\{ \mathbf{v}^\prime_1, \mathbf{v}^\prime_2 \}$ denote the precollisional
velocities of the spheres (the index `$1$' refers here  to species  $\alpha$),
and $\{ \mathbf{v}_1, \mathbf{v}_2 \}$ are 
the corresponding postcollisional velocities; 
$\hat{k}$ is a unit vector
pointing from the center of sphere $\alpha $ to that of sphere $\beta $, and
$M^{\alpha\beta}\equiv \frac{m_{\alpha }}{m_\alpha+m_\beta}$, 
 $\mathbf{v}_{12} \equiv 
\mathbf{v}_{1}-\mathbf{v}_{2}$, a similar definition holding for the
primed (precollisional) velocities. Obvious kinematic constraints require
that $\mathbf{v}^\prime_{12}\cdot \hat{k} \geq 0$. 
Define the degrees of inelasticity as: 
$\varepsilon_{\alpha\beta}\equiv 1-e_{\alpha\beta}^{2}$.

The kinetic description of a binary granular gas mixture involves two 
Boltzmann equations, one  for each 
species \cite{Garzo02,Serero06}: 
\begin {equation}
Df_{\alpha} \equiv 
\frac{\partial f_{\alpha }}{\partial t}+\mathbf{v}_{1}\cdot
\mathbf{\nabla }f_{\alpha }=\mathcal{B}(f_{\alpha},f_{\alpha},e_{\alpha \alpha})+\mathcal{B}(f_{\alpha},f_{\beta},e_{\alpha \beta})\ \ \ \ \  \alpha \ne \beta,
\label{Boltzmann}
\end{equation}
where  $f_{\alpha }\left( \mathbf{v}\right) $ is the 
distribution function for particles of species $\alpha$, and
\begin{displaymath}
 \mathcal{B}(f_{\alpha},f_{\beta},e_{\alpha \beta})
\end{displaymath}
\begin{equation}
=\sigma _{\alpha \beta }^{2}\int \int_{\mathbf{v}_{12}\cdot
\mathbf{k}>0}\left[ \frac{f_{\alpha }(\mathbf{v}_{1}^{\prime
})f_{\beta }(\mathbf{v}_{2}^{\prime })}{e_{\alpha \beta }^{2}}
-f_{\alpha }(\mathbf{v}_{1})f_{\beta }(\mathbf{v}_{2})\right]
\bigg( \mathbf{v}_{12}\cdot\mathbf{k}\bigg) \ d\mathbf{
v}_{2}\ d\mathbf{k},
\label{Boltzmannop}
\end{equation}
where $\sigma _{\alpha \beta }\equiv \frac{\sigma _{\alpha }+\sigma _{\beta }}{2}$. Here $\mathbf{v}_{1}$ and $\mathbf{v}_{1}^{\prime}$ pertain to species ${\alpha}$, and $\mathbf{v}_{2}$ and $\mathbf{v}_{2}^{\prime}$ pertain to $\beta$. Notice that $\mathcal{B}(f_{\alpha},f_{\beta},e_{\alpha \beta})$ depends  on the coefficients of normal restitution both explicitly, as
shown in Eq. (\ref{Boltzmannop}), and implicitly through the 
collision law.

The hydrodynamic fields
in the present case comprise the 
two number densities, $n_A$ and $n_B$ (or the  mass densities $\rho_A=n_Am_A$ and $\rho_B=n_Bm_B$), the mixture's 
 velocity field, $\mathbf{V}$, and the temperature field, $T$,  defined 
(differently from the monodisperse case) as twice  the 
mean fluctuating  kinetic energy of a particle (these are the only slow fields in the elastic limit), as explicitly defined below.
The continuum equations of motion 
 follow from the pertinent Boltzmann equation directly 
(using the  standard
procedure of computing velocity moments of the equation). Their validity
is general  since they are based on the underlying
conservation laws.
 The equation of motion for the number density, $n_\alpha$, 
with $\alpha \in \{ A,B\}$,  is:
\begin{equation}
\frac{Dn_{\alpha}}{Dt}=-{\mbox{div\ }} \mathbf{J}_{\alpha} -n_{\alpha}~{\mbox{div\ }}%
\mathbf{V}  \label{dynameq_na}
\end{equation}
where $\frac{D}{Dt}$ is the material derivative and 
%\begin{equation}
$\mathbf{J}_{\alpha}=n_{\alpha}\left( \mathbf{V}_{\alpha}-\mathbf{V}%
\right)  \label{field_ja}$
%\end{equation}
is the particle flux density of species $\alpha$. As ${\mathbf{V}}_\alpha$, 
the velocity field of
species $\alpha$ (or the flux  ${\mathbf{J}}_{\alpha}$) is not
a hydrodynamic field, it must be given by an appropriate constitutive
relation. 
The velocity field obeys (as expected):
%\begin{equation}
$\rho \frac{DV_{i}}{Dt}=-\frac{\partial \sigma_{ij}}{\partial x_{j}}+\rho \mathbf{g}$, 
%\label{dynameanveloc}
%\end{equation}
where the summation convention is used, 
and  $\sigma_{ij}$ is the stress tensor. 
The granular temperature field obeys:
%\begin{equation}
$n\frac{DT}{Dt}=T~{\mbox{div\ }}~\mathbf{J}-{\mbox{div\ }}~\mathbf{Q}
-2\sigma_{ij}\frac{\partial V_i}{{\partial x_j}}-\Gamma$,
%  \label{dynatemp}
%\end{equation}
where ${\mathbf{J}} \equiv {\mathbf{J}}_A + {\mathbf{J}}_B$ is the total
particle flux, and  $\mathbf{Q}$ is the heat flux.

The form of the constitutive relations can be easily determined
from tensorial (and symmetry) considerations, the result being (to 
linear order in the gradients, or Navier-Stokes order):
$
\sigma_{ij}=p\delta _{ij}-2\mu~D_{ij} -\delta_{ij} \ \eta_{\mbox{\tiny{B}}}~{\mbox{div\ }} \mathbf{V}$,
where 
%\begin{equation}
$D_{ij}=\frac{1}{2}\ \bigg( \frac{\partial V_{i}}{\partial x_{j}}+\frac{
\partial V_{j}}{\partial x_{i}} -\frac{2}{3}\  \delta_{ij} \ \mbox{div}{\bf V}\bigg)$
is the traceless rate of strain tensor, $\mu$ is the shear viscosity, 
$\eta_{\mbox{\tiny{B}}}$ is the 
bulk viscosity (which vanishes in the dilute limit) and $p$ is the
pressure. The diffusion flux can be rewritten as 
\begin{equation}
\mathbf{J}_{\alpha}=\frac{n_{\alpha}}{n}\frac{1}{\sigma_{AB}^2}\sqrt{\frac{T}{m_{\alpha}}}
\bigg(-\kappa_{\alpha}^{T}\mathbf{\nabla}\ln T -\kappa_{\alpha}^{n}\mathbf{\nabla}\ln n-\kappa_{\alpha}^{c}\mathbf{\nabla}\ln c \bigg),
\label{cr_jalpha2} 
\end{equation}
where $\kappa_\alpha^T$, $\kappa_\alpha^{n}$ and $\kappa_\alpha^{c}$
are non-trivial functions of the parameters
$${\mathcal{S}} \equiv \{ \{ \varepsilon_{\alpha\beta} \}, \  
M_A\equiv \frac{m_A}{m_A+m_B},\ \frac{\sigma_A}{\sigma_A + \sigma_B},\  c \},  $$
where the concentration field, $c$,  is defined as $c\equiv \frac{n_A}{n}
=\frac{n_A}{n_A+n_B}$. 
Similarly, the heat flux can be rewritten as
\begin{equation}
\mathbf{Q}=\frac{1}{\sigma_{AB}^2}\frac{T^{3/2}}{\sqrt{m_{0}}}\bigg( -\lambda^{T} \mathbf{\nabla}\ln T
-\lambda^{n}\mathbf{\nabla}\ln n-\lambda^{c}\mathbf{\nabla}\ln c \bigg)
\label {cr_q4}
\end{equation}
where $m_0 \equiv m_A + m_B$. In the dilute limit  the 
 equation of state  is the same as that for an
ideal gas: $p = \frac{nT}{3}$. 
Much like in the monodisperse case
 the number density for species $\alpha$ is given by:
%\begin{equation}
$n_{\alpha}=\int f_{\alpha} ( \mathbf{v}) \ d\mathbf{v}  $,
%\label{density}
%\end{equation}
the corresponding  mass density  being 
$\rho _{\alpha}=m_{\alpha}n_{\alpha}$; the overall number density is $n\equiv n_{A}+n_{B}$, and the overall mass density is $\rho\equiv \rho_{A}+\rho_{B}$.
The  velocity field of species $\alpha$ is given by:
%\begin{equation}
$\mathbf{V}_{\alpha}=\frac{1}{n_{\alpha}}\int f_{\alpha}
(\mathbf{v})\  \mathbf{v}
\ d\mathbf{v}  $.
%\label{speciesvelocity}
%\end{equation}
As mentioned,  $\mathbf{V}_\alpha$ is not a hydrodynamic field,  and  needs
to be expressed as a functional of the hydrodynamic fields. 
The mixture's  velocity field is:
$\mathbf{V}=\frac{1}{\rho }\left( \rho _{A}\mathbf{V}_{A}+\rho _{B}\mathbf{V}
_{B}\right)$. The  granular temperature of species $\alpha$ is defined by:
%\begin{equation}
$T_{\alpha}=\frac{1}{n_{\alpha}}\int f_{\alpha}(
 \mathbf{v})\  m_{\alpha}\ (\mathbf{v}-\mathbf{V})^{2}\ d
\mathbf{v}\  $
% \label{specitemp}
%\end{equation}
(not a hydrodynamic field).
The velocity fluctuations of each  of the species 
are measured with respect to the (hydrodynamic)
mixture's velocity field, not the species' velocity fields. 
The mixture's granular temperature 
is defined as:
$T=\frac{1}{n}\left( n_{A}T_{A}+n_{B}T_{B}\right)$.
The kinetic expression for the stress tensor \cite{Chapman70} is: 
%\begin{equation} 
$\sigma_{ij}=m_{A}\int f_{A}\left( \mathbf{v}\right) u_{i}u_{j}~d{\bf v}+m_{B}\int f_{B}\left( \mathbf{v}\right) u_{i}u_{j}~d{\bf v}$,
%\label{stresstensor}
%\end{equation}
where ${\mathbf{u}} \equiv \mathbf{v} - {\mathbf{V}}$ is the peculiar
(fluctuating) velocity of a particle (irrespective of the species). 
Similarly, the heat flux is composed of two contributions, for obvious reasons:
%\begin{equation}
$\mathbf{Q}=\int f_{A}\left( \mathbf{v}\right) m_{A}u^{2}
\mathbf{u}\ d{\bf v} +\int f_{B}\left( \mathbf{v}\right) m_{B}u^{2}
\mathbf{u}\ d{\bf v}$.
%\label{heatflux}
%\end{equation}
 Our definition of the heat flux 
may differ by a factor of $2$ from some other definitions.
The   sink term is given by $\Gamma = \Gamma_A + \Gamma_B + \Gamma_{AB}$,
where:
\begin{equation}
\Gamma _{\alpha} \equiv \varepsilon_{\alpha\alpha} \frac{ m_{\alpha}
     \pi\sigma_{\alpha}^{2}}{8}
\int \int f_{\alpha}\left( \mathbf{v}_{1}\right) f_{\alpha}\left( 
\mathbf{v}_{2}\right) \left| v_{12}\right| ^{3}d{\bf v}_{1}\ d{\bf v}_{2},
\label {GammaA}
\end{equation}
and 
\begin{equation}
\Gamma _{AB}\equiv \varepsilon _{AB}\frac{m_{AB}\pi \sigma_{AB}^{2}}{2}\int \int
f_{A}\left( \mathbf{v}_{1}\right) f_{B}\left (\mathbf{v}_{2}\right) \left| \mathbf{v}%
_{12}\right| ^{3}d{\bf v}_{1}\ d{\bf v}_{2}.  \label{GammaAB}
\end{equation}

The HCS distributions for binary granular gas mixtures (much
like in the  monodisperse case) are 
scaling solutions, in which the distribution functions are rendered  time independent 
when the velocities are 
 scaled by the respective (decaying)  thermal speeds of the species comprising the 
mixture. It serves as a zeroth order in the extended Chapman-Enskog
expansion of \cite{Garzo02}.  
The lhs of the Boltzmann equations for the HCS can be rewritten in this case as: %\begin{equation}
$\frac{\partial f^{\mbox{\tiny{HCS}}}_{\alpha }}{\partial t}=\frac{\partial f^{\mbox{\tiny{HCS}}}_{\alpha }}{\partial
T}\frac{\partial T}{\partial t}=-\frac{\Gamma }{n}\frac{\partial f^{\mbox{\tiny{HCS}}}_{\alpha }
}{\partial T}$
%\end{equation}
and the two Boltzmann equations become: 
\begin{equation}
-\frac{\Gamma }{n}\frac{\partial f^{\mbox{\tiny{HCS}}}_{\alpha}}{\partial T}=
\mathcal{B}\left(
f^{\mbox{\tiny{HCS}}}_{\alpha},f^{\mbox{\tiny{HCS}}}_{\alpha,}e_{\alpha\alpha}\right) +\mathcal{B}\left( f^{\mbox{\tiny{HCS}}}_{\alpha},f^{\mbox{\tiny{HCS}}}_{\beta,}
e_{\alpha\beta}\right) \ \ \ \ \ \ \ \ \alpha \ne \beta \label{boltHCSA}  
\end{equation}
These equations need to be solved subject to the following  `constraints'
which fix the values of the hydrodynamic fields: 
$\int f^{\mbox{\tiny{HCS}}}_{A}d\mathbf{u} =n_{A}$, $\int f^{\mbox{\tiny{HCS}}}_{B}d\mathbf{u} =n_{B}$ and
$\int f^{\mbox{\tiny{HCS}}}_{A}m_{A}u^{2}d\mathbf{u+}\int f^{\mbox{\tiny{HCS}}}_{B}m_{B}u^{2}d\mathbf{u} =nT$.
The common way of solving Eqs. (\ref{boltHCSA}) is to represent 
the functions, $\ f^{\mbox{\tiny{HCS}}}_{\alpha }$,
by   Sonine polynomial series times  Maxwellians in the respective  non-dimensionalized 
(peculiar) velocities.  However, it can be shown \cite{Noskowicz06} that due to the exponential 
tails  of the HCS distributions these expansions are not convergent 
yet they are asymptotic and Borel resummable. To overcome this difficulty,  consider the modified expansion:   
\begin{equation}
f^{\mbox{\tiny{HCS}}}_{\alpha }=f_{\alpha}^{M,\eta}\phi
_{\alpha } \equiv n_{\alpha }\left( \frac{\gamma
_{\alpha }}{\pi }\right) ^{\frac{3}{2}}e^{-\eta \gamma _{\alpha }u^{2}}\phi
_{\alpha } \label{expansion}
\end{equation}
where $\gamma _{\alpha }=\frac{3m_{\alpha }}{2T}$,  $\phi _{\alpha
}\equiv\sum_{p=0}^{\infty }h_{\alpha }^{p}S_{\frac{1}{2}}^{p}\left(
\gamma _{\alpha }u^{2}\right) $, and $\eta>0$ is a constant, i.e. $f_{\alpha }$ is represented by
a truncated Sonine polynomial series (whose coefficients are denoted
by $ \{ h^p_\alpha \} $) times a Maxwellian in which  the
temperature is replaced by a ``wrong'' temperature $\frac{T}{\eta }$.  It
can be shown \cite{Noskowicz06} 
that this method produces convergent series when  $\eta
<0.5$.
Upon substituting the form (\ref{expansion}) in Eq. (\ref{boltHCSA}) and projecting on the Nth order Sonine polynomial  $S_{\frac{1}{2}}^{N}\left( \gamma _{\alpha }u_{1}^{2}\right) $, one obtains the following non linear algebraic system for the coefficients $h_{\alpha}^p$:
\begin{eqnarray}
\frac{n_{\alpha}\Gamma }{nT}\sum\limits_{p}^{ }h_{\alpha}^{p}R_{pN}
 &=&\frac{M_{\alpha }^{3}n_{\alpha }^{2}\sigma _{\alpha }^{2}}{
\pi ^{3}}\sqrt{\frac{2T}{3m_{0}}}\sum\limits_{p,q}B_{\alpha
\alpha }^{pqN}h_{\alpha }^{p}h_{\alpha }^{q} \nonumber\\
&+&\chi _{\alpha \beta }\frac{\left( M_{\beta }M_{\alpha }\right) ^{\frac{3}{
2}}n_{\alpha }n_{\beta }\sigma _{\alpha \beta }^{2}}{\pi ^{3}}\sqrt{\frac{2T
}{3m_{0}}}\sum\limits_{p,q} B_{\alpha \beta }^{pqN}h_{\alpha
}^{p}h_{\beta }^{q}\label{Bolteqexp}
\end{eqnarray}
where $m_0 \equiv m_A+m_B$, $M_{\alpha} \equiv \frac{m_{\alpha}}{m_0}$, and  $R_{pN}$ is defined as follows:
\begin{displaymath}
R_{pN}=\frac{1}{\pi ^{\frac{3}{2}}}\int e^{-\eta u_{\alpha}^{2}}S_{\frac{1}{2}
}^{N}\left( u_{\alpha}^{2}\right)
\end{displaymath}
\begin{equation}
\times
\left[ \frac{3}{2}S_{\frac{1}{2}}^{p}\left(
u_{\alpha}^{2}\right) -u^{2}\left( \eta S_{\frac{1}{2}}^{p}\left( u_{\alpha}^{2}\right) -\frac{
\partial }{\partial u_{\alpha}^{2}}S_{\frac{1}{2}}^{p}\left( u_{\alpha}^{2}\right)
\right) \right] d\mathbf{u_{\alpha}} \label{int_R}
\end{equation}
where $\mathbf{u}_{\alpha}\equiv \sqrt{\gamma_{\alpha}}\mathbf{u}$, and the coefficients (or coupling constants),  $B_{\alpha \beta }^{pqN}$, are defined, in terms of the rescaled velocities $\tilde{\mathbf{u}} \equiv \sqrt{\frac{3m_0}{2T}}\mathbf{u}$, by:
\begin{eqnarray}
B_{\alpha \beta }^{pqN} &= &\int S_{\frac{1}{2}}^{N}\left( M _{\alpha }\tilde{u}_{1}^{2}\right)\nonumber \int \int_{\mathbf{\tilde{u}}_{12}\mathbf{\cdot k}>0}\bigg[\frac{1}{e_{\alpha \beta }^{2}}
e^{-\eta M _{\alpha }\tilde{u}_{1}^{\prime 2}-\eta M _{\beta
}\tilde{u}_{2}^{\prime 2}} S_{\frac{1}{2}}^{p}\left( M _{\alpha }\tilde{u}_{1}^{\prime 2}\right)
S_{\frac{1}{2}}^{q}\left( M _{\beta }\tilde{u}_{2}^{\prime 2}\right)\nonumber\\
& & -e^{-\eta
M _{\alpha }\tilde{u}_{1}^{2}-\eta M _{\beta }\tilde{u}_{2}^{2}}S_{\frac{1}{2}}^{p}\left( M _{\alpha }\tilde{u}_{1}^{2}\right) S_{
\frac{1}{2}}^{q}\left( M _{\beta }\tilde{u}_{2}^{2}\right)\bigg] \left( \mathbf{\tilde{u}}
_{12}\mathbf{\cdot k}\right) d\mathbf{\tilde{u}}_{1}d\mathbf{\tilde{u}}_{2}d\mathbf{k}
\label{int_B}
\end{eqnarray}
Though straightforward in principle, the computation of these coefficients
turns out to be forbiddingly tedious as  the order of the Sonine expansions is
increased. In order to be able to carry out high order expansions, we made use of a computer-aided method exploiting  the fact that 
the Sonine polynomials can be derived from their respective
 generating functions,
$G_m \left(s,x\right)$:
\begin {equation}
G_m\left( s,x\right)=\left( 1-s\right)
^{-m-1}e^{-\frac{s}{1-s}x} =
\sum\limits_{p=0}^\infty s^{p}S_{m}^{p}\left( x\right)
\label{Sonine_gf}
\end{equation}
This fact enables one to define generating functions for the
integrals (\ref{int_R}) and (\ref{int_B}):  $R\left( s,w\right)
\equiv \sum\limits_{p,N}s^{p}w^{N}R_{pN}$ and
$B_{\alpha
\beta }\left( w,s,t\right)
\equiv\sum\limits_{p,q,N}s^{p}t^{q}w^{N}B_{\alpha \beta }^{pqN}$
of
\ the  coefficients, $B_{\alpha \beta }^{pqN}$, and $R_{pN}$,
respectively (where all  sums here and below range from zero to infinity). The latter can then be computed (using a symbolic processor
such as MAPLE$^{\mbox{\tiny{TM}}}$) to any desired order by
taking successive derivatives of  $R\left( s,w\right) $ and  $B_{\alpha
\beta }\left( w,s,t\right)$.
%\begin{equation}
%R_{pN} = \frac{1}{p!N!}\frac{\partial^{p+N}R\left( w,s\right)}{\partial s^{p}\partial w^{N}}\rfloor s%=0,w=0}
%\end{equation}
%and
%\begin{equation}
%B_{\alpha \beta }^{pqN} = \frac{1}{p!q!N!}\frac{\partial^{p+q+N}B_{\alpha
%\beta }\left( w,s,t\right)}{\partial s^{p}\partial t^{q} \partial w^{N}}\rfloor _{s=0,t=0,w=0}
%\end{equation}
%where: 
These generating functions are given by: 
\begin{equation}
R\left( s,w\right) =\frac{3}{2}\frac{\left( 1-s\right) w}{\left( 1-ws+\left(
\eta -1\right) \left( 1-w\right) \left( 1-s\right) \right) ^{\frac{5}{2}}}
\label{geneR}
\end{equation}
and using Eq. (\ref{int_B})
\begin{eqnarray}
B_{\alpha \beta } &=&\left( 1-s\right) ^{-\frac{3}{2}
}\left( 1-t\right) ^{-\frac{3}{2}}\left( 1-w\right) ^{-\frac{3}{2}}\nonumber \\
&&\times \bigg[ \frac{1}{e_{\alpha \beta }^{2}} I_{\alpha \beta }\left( 
\frac{M_{\alpha }w}{1-w},0,\left( \frac{s}{1-s}+\eta \right) M_{\alpha
},\left( \frac{t}{1-t}+\eta \right) M_{\beta }\right)  \nonumber\\
&&\qquad-I_{\alpha \beta }\left( M_{\alpha }\left( \eta +\frac{w}{1-w}+\frac{s}{1-s
}\right) ,\left( \frac{t}{1-t}+\eta \right) M_{\beta }\right) \bigg]\nonumber
\end{eqnarray}
where $I_{\alpha \beta }\left( a,b,c,d,x,y,z\right) $ is the integral defined by:
\begin{equation}
I_{\alpha \beta }\left( a,b,c,d,x,y,z\right) \equiv \int_{\mathbf{u}_{12}
\mathbf{\cdot k}>0}d\mathbf{u}_{1}d\mathbf{u}_{2}d\mathbf{k}\left( 
\mathbf{u}_{12} \cdot \mathbf{
k} \right) e^{-F}
\label{IAB}
\end{equation}
where 
$
F=au_{1}^{2}+bu_{2}^{2}+cu_{1}^{\prime 2}+du_{2}^{\prime 2}+\frac{x}{2}%
\left( \mathbf{u}_{1}+\mathbf{u}_{1}^{\prime }\right) ^{2}+\frac{y}{2}\left( 
\mathbf{u}_{1}\mathbf{+u}_{2}^{\prime }\right) ^{2}+\frac{z}{2}\left( 
\mathbf{u}_{1}\mathbf{+u}_{2}\right) ^{2}
$ and $I_{\alpha\beta}\left( a, b \right) \equiv I_{\alpha\beta}\left(a,b,0,0
\right)$. 
It is easy to calculate $I_{\alpha \beta } $, the result being:
\begin{equation}
I_{\alpha \beta }=\frac{2\pi ^{\frac{7}{2}}}{\lambda ^{\frac{3}{2}}}\frac{1}{
\mu _{\alpha \beta }\left( \nu _{\alpha \beta }+\mu _{\alpha \beta }\right) }
\end{equation}
where $\lambda=a+b+c+d+2x+2y+2z$, $\mu _{\alpha \beta }=R_{\alpha \beta }-\frac{K_{\alpha \beta }^{2}}{\lambda },\ \ 
 \nu _{\alpha \beta }=S_{\alpha \beta }$\\
$$-\frac{t^{\alpha\beta}}{\lambda }\left(
\left( d\mathbf{+}y\right) M^{\alpha \beta }-\left( c+x\right) M^{\beta
\alpha }\right) \left( 2K_{\alpha\beta}+t^{\alpha\beta}\left( \left( d\mathbf{+}y\right) M^{\alpha \beta
}-\left( c+x\right) M^{\beta \alpha }\right) \right), $$ \\
$R_{\alpha \beta } =\left( a+c+2x\right) \left( M^{\beta \alpha }\right)
^{2}+\left( b+d\right) \left( M^{\alpha \beta }\right) ^{2}+\frac{1}{2}%
\left( M^{\alpha \beta }-M^{\beta \alpha }\right) ^{2}\left( y+z\right)$, \\
$S_{\alpha \beta } =\frac{1}{2}t^{\alpha\beta}\left[ \left( t^{\alpha\beta}-2\right) \left( \left(
2c+x\right) \left( M^{\beta \alpha }\right) ^{2}+\left( M^{\alpha \beta
}\right) ^{2}\left( 2d+y\right) \right) +2M^{\alpha \beta } M^{\beta\alpha} y 
\right. $$\\
$$\left. 
-2\left(
M^{\beta \alpha }\right) ^{2}x\right] $, 
$K_{\alpha \beta }=\left( M^{\beta \alpha }\left( a+c+2x+y+z\right)
-M^{\alpha \beta }\left( b+d+y+z\right) \right) $, and
$t^{\alpha\beta}\equiv \frac{1+e_{\alpha\beta}}{e_{\alpha\beta}}$.
Next,  using Eq.~(\ref{expansion}) for $f_{\alpha }$ in Eq. (\ref{GammaA}) and (\ref{GammaAB}), multiplying the result by $s^pt^q$ and summing over $p$ and $q$, one obtains:
$\Gamma =\frac{\sqrt{6\pi }}{9}\frac{n^{2}\sigma _{AB}^{2}T^{\frac{3}{2}}}{
\sqrt{m_{0}}}\widetilde{\Gamma }
$, where
\begin{eqnarray}
\widetilde{\Gamma } &\equiv&\varepsilon _{AA}\frac{n_{A}^{2}}{n^{2}}
\frac{\sigma _{A}^{2}}{\sigma _{AB}^{2}}M_{A}^{4}\sum_{p}\sum_{p}\Gamma _{AA}^{pq}h_{A}^{p}h_{A}^{q} 
%\nonumber \\&&
+4\varepsilon _{AB}\frac{n_{A}n_{B}}{n^{2}}M_{A}^{\frac{5}{2}
}M_{B}^{\frac{5}{2}}\sum_{p}\sum_{q}\Gamma
_{AB}^{pq}h_{A}^{p}h_{B}^{q} \nonumber \\
&&+\chi _{BB}\varepsilon _{BB}\frac{n_{B}^{2}}{n^{2}}\frac{\sigma _{B}^{2}}{
\sigma _{AB}^{2}}M_{B}^{4}\sum_{p}\sum_{q}\Gamma
_{BB}^{pq}h_{B}^{p}h_{B}^{q},
\label{gammaexp}
\end{eqnarray}
$\Gamma _{\alpha \beta }^{pq} \equiv \frac{1}{p!q!}\frac{\partial ^{p+q}G_{\alpha
\beta }^{\Gamma }}{\partial s^{p}\partial t^{q}}\rfloor _{s=0,t=0}
$, 
the generating function $G_{\alpha \beta }^{\Gamma }\left( s,t\right) $ being given by:
\begin{equation}
G_{\alpha \beta }^{\Gamma }\left( s,t\right) =\frac{\left( \frac{\left( 1-t\right) M_{\alpha }}{\left( t+\eta \left(
1-t\right) \right) }+\frac{\left( 1-s\right) M_{\beta }}{\left( s+\eta
\left( 1-s\right) \right) }\right) ^{\frac{3}{2}}}{\left( s+\eta \left(
1-s\right) \right) ^{\frac{3}{2}}\left( t+\left( 1-t\right) \eta \right) ^{
\frac{3}{2}}M_{\alpha }^{3}M_{\beta }^{3}}
\label{genegamma}
\end{equation}
%Finally, using the form (\ref{expansion})  the aforementioned
%kinematic constraints  read:
%\begin{equation}
%\sum\limits_{p=0}^{\infty }O_{p}^{I}h_{A}^{p}=1;\ \ \sum\limits_{p=0}^{\infty }O_{p}^{I}h_{B}^{p}=1; %\ \
%\sum\limits_{p=0}^{\infty }O_{p}^{II}\left( \frac{n_{A}}{n}h_{A}^{p}+\frac{
%n_{B}}{n}h_{B}^{p}\right) =1
%\end{equation}
%where the coefficients $O_{p}^{I}$ and $O_{p}^{II}$ are obtained using their
%respective generating functions: 
%\begin{equation}
%O^{I}\left( s\right)  \equiv \sum\limits_{p=0}^{\infty }s^{p}O_{p}^{I}=\frac{1}{\left( s+\left( 1-s\r%ight) \eta
%\right) ^{\frac{3}{2}}}; \ 
%O^{II}\left( s\right)  \equiv \sum\limits_{p=0}^{\infty }s^{p}O_{p}^{II}=\frac{\left( 1-s\right) }{\l%eft( s+\left(
%1-s\right) \eta \right) ^{\frac{5}{2}}}
%\end{equation}

\begin{figure}
\begin{center}
\begin{tabular}{ccc}
\subfigure[]{\includegraphics[height=0.25\hsize]{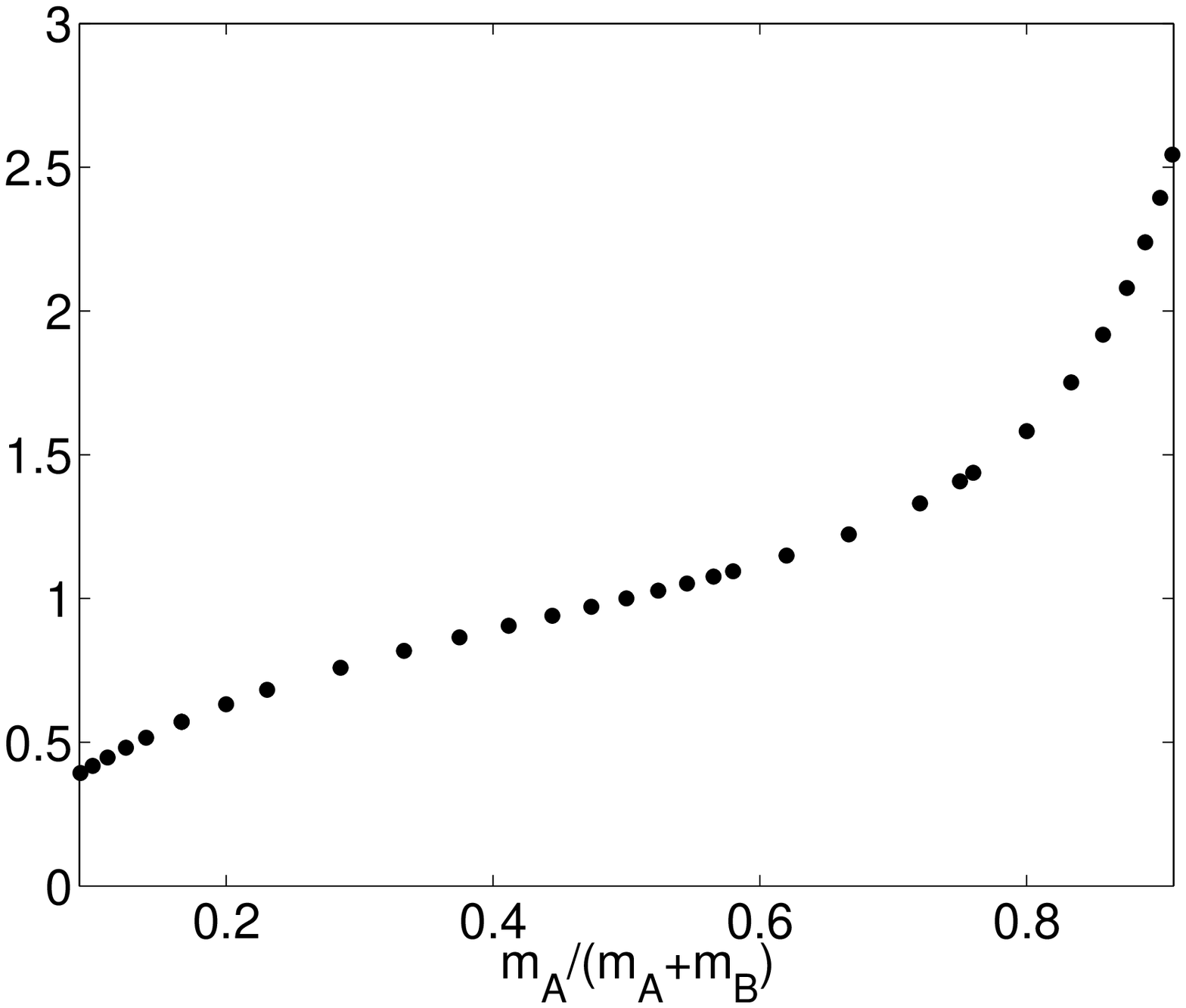}}&
\subfigure[]{\includegraphics[height=0.25\hsize]{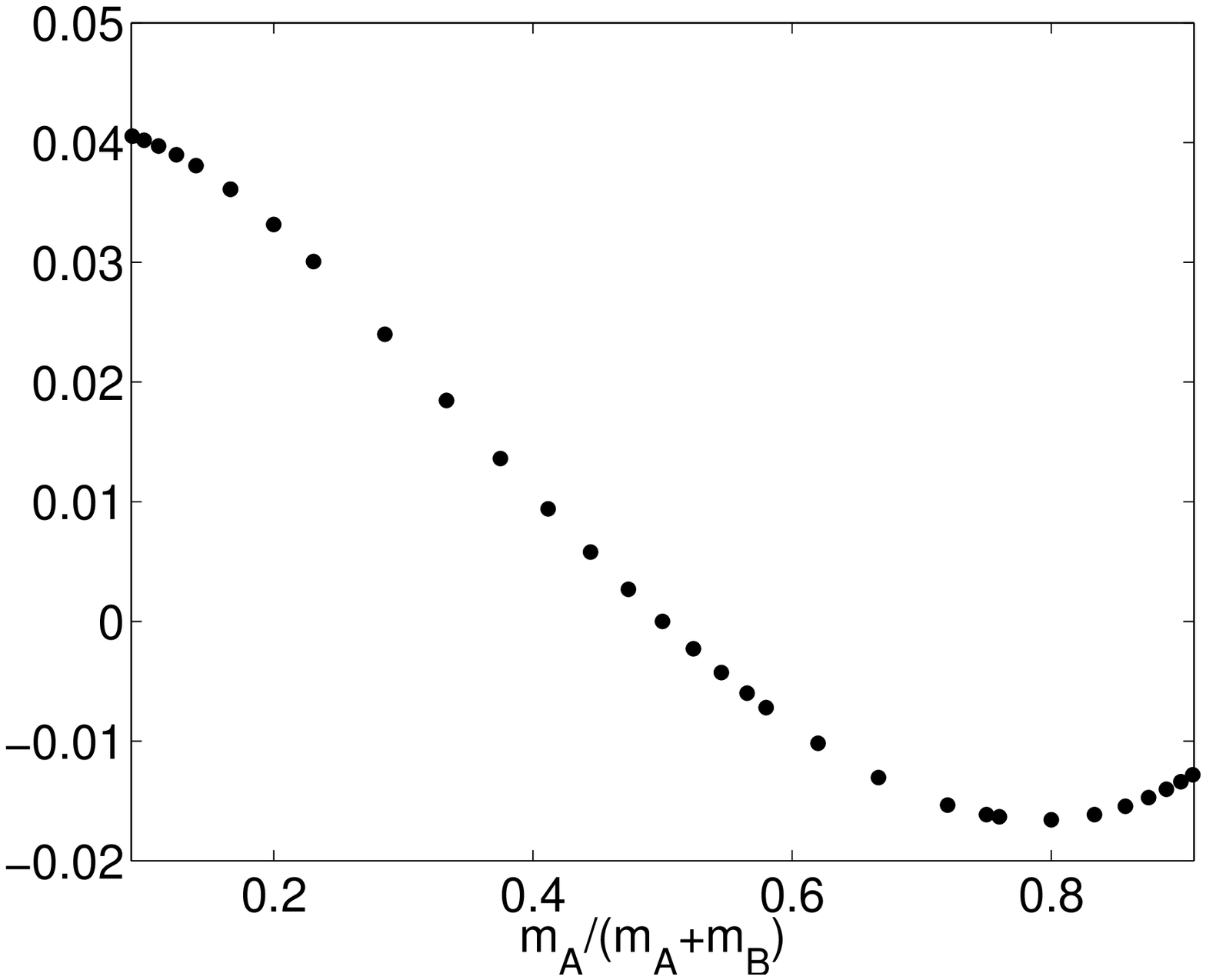}}&
\subfigure[]{\includegraphics[height=0.25\hsize]{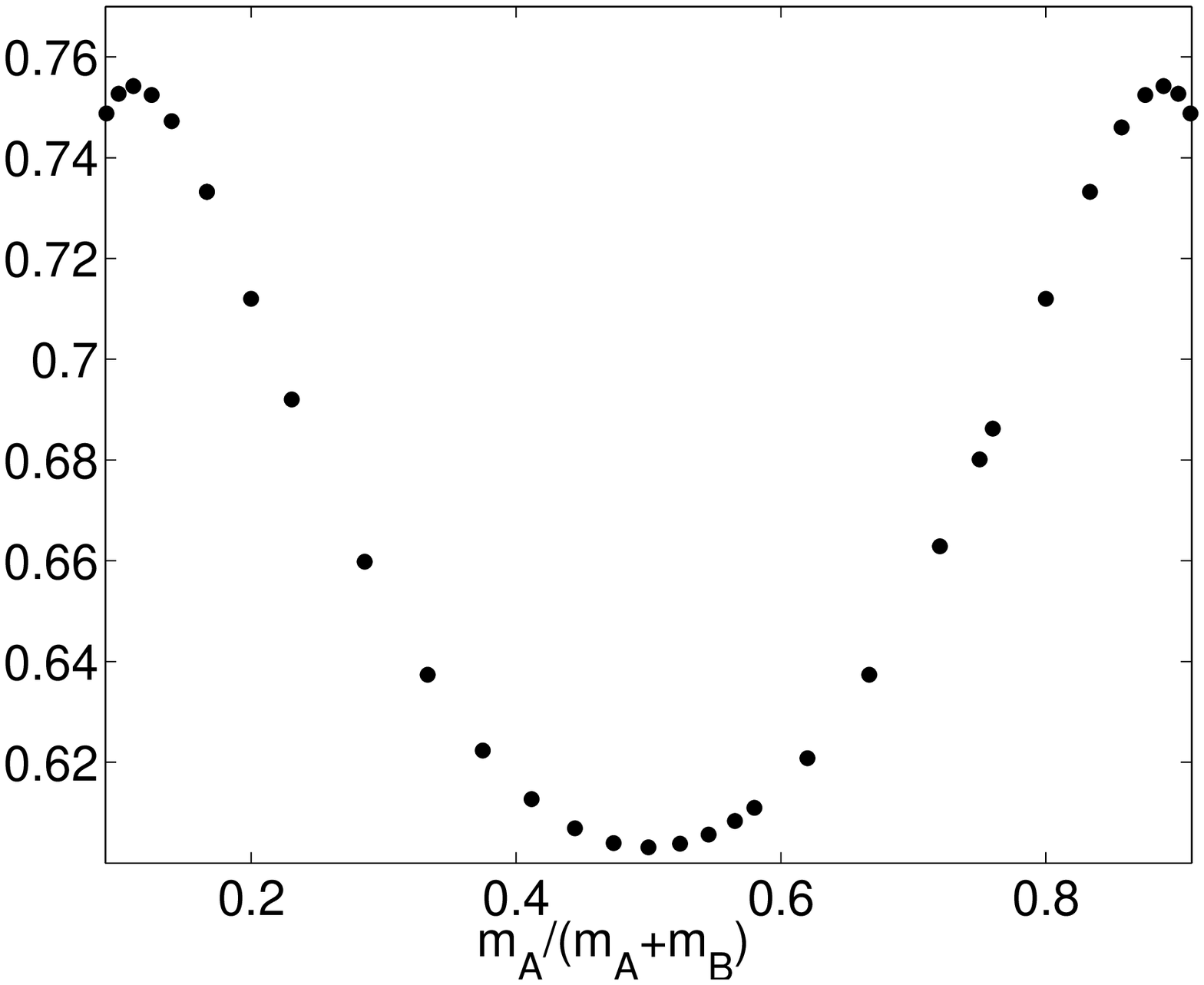}}
\end{tabular}
\end{center}
\caption{Plot of the  temperature ratio $T_A/T_B$  (a), thermal diffusion coefficient $\kappa_A^{T}$ (b), and thermal
  conductivity $\lambda^{T}$  (c), as a
  function of $\frac{m_A}{m_A+m_B}$ for grains made of the same material, with
  $\frac{n_A}{n}=\frac{n_B}{n}=0.5$ and
  $\varepsilon_{AA}=\varepsilon_{AB}=\varepsilon_{BB}=0.8$}
\label{TR_fig}
\end{figure}

%As mentioned, a truncation of the above equations can be solved for any values
%of the parameters of the problem, by using standard solvers. Convergence is
%tested  by checking that an increase in the order of truncation (we have
%checked up to the 25th order in the Sonine polynomial expansion in some cases)
%does not change the results (i.e., the desired coefficients). 
With the coupling constants calculated,  the algebraic equations, Eq.~(\ref{Bolteqexp}),  are now solved to produce the distribution
functions, from which one can easily calculate the temperature ratio,
$T_A/T_B$. 
Fig.~\ref{TR_fig}(a) shows a plot of the temperature ratio as a function of 
$\frac{m_A}{m_A +m_B}$ for a system of particles of equal mass density, $c=0.5$ and $\varepsilon_{AA}=\varepsilon_{BB}=\varepsilon_{AB}=0.8$. Note that the temperature ratio can ``violate'' equipartition quite strongly.
In order to compute the heat conductivity and the thermal diffusion coefficient, we now turn to solve the Boltzmann equation, carrying out the Chapman-Enskog expansion to first order in gradients of the hydrodynamic fields, with the homogeneous cooling state distribution function serving as as a zeroth order solution. The distribution functions are expressed as $f_{\alpha }=f_{\alpha
}^{\mbox{\tiny{HCS}}}+f_{\alpha }^{K}$, where $f_{\alpha }^{K}$ is the first order
perturbation in a gradient expansion, written for convenience as $
f_{A}^{K}=f_{\alpha }^{M,\nu }\Phi _{\alpha }^{K},$ cf. Eq.~(\ref{expansion}), where the value of $\nu$ may be taken to be different from that of $\eta$.  To first order in the gradients of
the hydrodynamic fields, the two coupled Boltzmann equations for the
perturbations  $\Phi _{\alpha }^{K}$ read:
\begin{eqnarray*}
D^{K}f_{A} &=&\left( L_{AA}^{(1)}+L_{AA}^{(2)}+L_{AB}^{(1)}\right) \Phi
_{A}^{K}+L_{AB}^{(2)}\Phi _{B}^{K} \\
D^{K}f_{B} &=&L_{BA}^{(2)}\Phi _{A}^{K}+\left(
L_{BB}^{(1)}+L_{BB}^{(2)}+L_{BA}^{(1)}\right) \Phi _{B}^{K}
\end{eqnarray*}
where 
\begin{eqnarray*}
L_{\alpha \beta }^{(1)}\Phi _{\alpha }^{K} &\equiv &\mathcal{B}\left(
f_{\alpha }^{M,\nu }\Phi _{\alpha }^{K},f_{\beta }^{\left( HCS\right)
},e_{\alpha \beta }\right)  \ \ \  \alpha\ne\beta\\
L_{\alpha \beta }^{(2)}\Phi _{\beta }^{K} &\equiv &\mathcal{B}\left(
f_{\alpha }^{\left( HCS\right) },f_{\beta }^{M,\nu }\Phi _{\beta
}^{K},e_{\alpha \beta }\right) \ \ \ \alpha \ne \beta
\end{eqnarray*} 
are the  linearized Boltzmann operators,  
and $D^{K}f_{\alpha }$ denotes the first order term in the expansion of $Df_{\alpha }$. Following the definition of the operator, $D$:
\begin{eqnarray*}
Df_{\alpha} &=&f_{\alpha }^{M,\eta }\bigg[ \phi_{\alpha} D\ln n_{\alpha}+2\gamma
_{\alpha}u_{i}\left( \eta \phi_{\alpha} -\phi_{\alpha} ^{\prime }\right) DV_{i}\\
&+&\left( \gamma
_{\alpha}u^{2}\left( \eta \phi_{\alpha} -\phi_{\alpha} ^{\prime }\right) -\frac{3}{2}\phi_{\alpha} \right)
D\ln T+c\frac{\partial \phi_{\alpha} }{\partial c}D\ln c\bigg]  \\
&+&f_{\alpha }^{M,\nu }D\Phi _{\alpha}^{K}+\Phi _{\alpha}^{K}f_{\alpha }^{M,\nu }\left[
D\ln n_{\alpha}+2\gamma _{\alpha}u_{i}\nu DV_{i}+\left( \gamma _{\alpha}u^{2}\nu -\frac{3}{2%
}\right) D\ln T\right] 
\end{eqnarray*}
where a prime superscript of a function denotes a derivative with respect
to its argument. 
Carrying out the expansion to first order in gradients leads, on the basis
of tensorial considerations, to the following form for the function $\Phi
_{\alpha}^{K}$:
\begin{eqnarray}
\Phi _{\alpha}^{K}&=&\Phi _{\alpha}^{K,T}\left( \gamma _{\alpha}u^{2}\right) \sqrt{\gamma _{\alpha}%
}u_{j}\frac{\partial \ln T}{\partial x_{j}}+\Phi _{\alpha}^{K,n}\left( \gamma
_{\alpha}u\right) \sqrt{\gamma _{\alpha}}u_{j}\frac{\partial \ln n}{\partial x_{j}} \nonumber\\
&+&\Phi _{\alpha}^{K,c}\left( \gamma _{\alpha}u\right) \sqrt{\gamma _{\alpha}}u_{j}\frac{%
\partial \ln c}{\partial x_{j}}+\Phi _{\alpha}^{K,V}\left( \gamma _{\alpha}u\right)
\gamma _{\alpha}^{\frac{3}{2}}\overline{u_{i}u_{j}}\frac{\partial V_{i}}{\partial
x_{j}} \label{fiKform}
\end{eqnarray}
where for any tensor $A_{ij}$, $\overline{A_{ij}}\equiv \frac{1}{2}\left(A_{ij}+A_{ji}-\frac{2}{3}\delta_{ij}A_{kk}\right)$. $D^{K}f_{\alpha }$ is given by: 
\begin{eqnarray}
D^{K}f_{\alpha } &=&\frac{\Gamma }{nT}f_{\alpha }^{M,\nu }\bigg[\left( \gamma
_{\alpha }u^{2}\left( \Phi _{\alpha }^{K,c}\right) ^{\prime }-\left( \gamma
_{\alpha }u^{2}\nu -2\right) \Phi _{\alpha }^{K,c}-c\frac{\partial \ln 
\widehat{\Gamma }}{\partial c}\Phi _{\alpha }^{K,T}\right) \nonumber\\
& & \times \sqrt{\gamma
_{\alpha }}u_{j}\frac{\partial \ln c}{\partial x_{j}} \nonumber\\
&&+\left( \gamma _{\alpha }u^{2}\left( \Phi _{\alpha }^{K,n}\right) ^{\prime
}-\left( \gamma _{\alpha }u^{2}\nu -2\right) \Phi _{\alpha }^{K,n}-\Phi
_{\alpha }^{K,T}\right) \sqrt{\gamma _{\alpha }}u_{j}\frac{\partial \ln n}{%
\partial x_{j}}\nonumber \\
&&+\left( \gamma _{\alpha }u^{2}\left( \Phi _{\alpha }^{K,T}\right) ^{\prime
}-\left( \gamma _{\alpha }u^{2}\nu -2\right) \Phi _{\alpha }^{K,T}-\frac{1}{2%
}\Phi _{\alpha }^{K,T}\right) \sqrt{\gamma _{\alpha }}u_{j}\frac{\partial
\ln T}{\partial x_{j}}\nonumber \\
&&+\left( \gamma _{\alpha }u^{2}\left( \Phi _{\alpha }^{K,V}\right) ^{\prime
}-\left( \gamma _{\alpha }u^{2}\nu -3\right) \Phi _{\alpha }^{K,V}\right)
\gamma _{\alpha }^{\frac{3}{2}}\overline{u_{i}u_{j}}\frac{\partial V_{i}}{%
\partial x_{j}}\bigg]  \nonumber\\
&&
+f_{\alpha }^{M,\eta }\bigg[2\gamma _{\alpha }\left( \eta \phi _{\alpha
}-\phi _{\alpha }^{\prime }\right) \overline{u_{i}u_{j}}\frac{\partial V_{i}%
}{\partial x_{j}}+\left( \phi _{\alpha }-\frac{nm_{\alpha}}{\rho }\left( \eta \phi _{\alpha
}-\phi _{\alpha }^{\prime }\right) \right) u_{i}\frac{\partial \ln n}{%
\partial x_{i}} \nonumber\\
&&+\left( \left( \gamma _{\alpha }u^{2}-\frac{nm_{\alpha }}{\rho }\right)
\left( \eta \phi _{\alpha }-\phi _{\alpha }^{\prime }\right) -\frac{3}{2}%
\phi _{\alpha }\right) u_{i}\frac{\partial \ln T}{\partial x_{i}} \nonumber\\
&&+f_{\alpha }^{M,\eta }\left( \phi _{\alpha }+c\frac{\partial \phi _{\alpha
}}{\partial c}\right) u_{j}\frac{\partial \ln c}{\partial x_{j}}\bigg]
\label{DKf}
\end{eqnarray}
%Employing the form (\ref{fiKform}) into the definitions of the heat and particle fluxes
%yields the following expressions for the heat conductivity and thermal diffusion coefficient%s: 
Employing (\ref{fiKform}) in the definition of $\mathbf{Q}$ and $\mathbf{J}_A$, one obtains:
\begin{eqnarray*}
\mathbf{\kappa }_{A}^{T} &=&-\frac{1}{3}Tr\left[ \int f_{A}^{M,\nu}\left[
\Phi _{A}^{K,T}\left( \gamma _{A}u^{2}\right) \sqrt{\gamma _{A}}u_{j}\right]
u_{i}d\mathbf{u}\right]  \\
\mathbf{\lambda }^{T} &=&-\frac{1}{3}Tr\left[ \int\left( m_{A}f_{A}^{M,\nu}\Phi
_{A}^{K,T}\left( \gamma _{A}u^{2}\right) \sqrt{\gamma _{A}}   \right. \right.\\
&& \left. \left. \qquad \qquad \ \  +m_{B} f_{B}^{M,\nu}\Phi _{B}^{K,T}\left( \gamma _{B}u^{2}\right) \sqrt{%
\gamma _{B}}\right)\right] u^{2}u_iu_jd\mathbf{u}
\end{eqnarray*}
where $\Phi _{\alpha }^{K,T}$ obey (with $\beta\ne\alpha$):
\begin{eqnarray}
&&\left( L_{\alpha \alpha }^{(1)}+L_{\alpha \alpha }^{(2)}+L_{\alpha \beta
}^{(1)}\right) \left( \Phi _{\alpha }^{K,T}\left( \gamma _{\alpha }u^{2}\right) \sqrt{\gamma _{\alpha }}\mathbf{u}%
\right) +L_{\alpha \beta }^{(2)}\left( \Phi _{\beta }^{K,T}\left( \gamma _{\alpha }u^{2}\right) \sqrt{\gamma
_{\beta }}\mathbf{u}\right) \nonumber\\
&-&\frac{\sqrt{6\pi }}{9}\frac{n\sigma _{\alpha
\beta }^{2}}{\sqrt{m_{0}}}\widetilde{\Gamma }\sqrt{T}\mathbf{H}_{\alpha
}^{T}\left( \Phi _{\alpha }^{K,T}\left( \gamma _{\alpha }u^{2}\right) \sqrt{\gamma _{\alpha }}\mathbf{u}\right) =%
\mathbf{R}_{\alpha }^{T}
\label{EqfiKT}
\end{eqnarray}

where\ $\mathbf{H}_{\alpha}^{T}\left( \Phi _{\alpha }^{K,T}\sqrt{\gamma _{\alpha }}
\,\mathbf{u}\right) =f_{\alpha }^{M,\nu }\left( \gamma _{\alpha}u^{2}\left( \Phi
_{\alpha}^{K,T}\right) ^{\prime }-\left( \gamma _{\alpha}u^{2}\nu -\frac{3}{2}\right)
\Phi _{\alpha}^{K,T}\right) \sqrt{\gamma _{\alpha}}\,\mathbf{u}\,$ and $\mathbf{R}
_{\alpha}^{T}=f_{\alpha }^{M,\nu }\left( \left( \gamma _{\alpha}u^{2}-\frac{nm_{\alpha}}{
\rho }\right) \left( \eta \phi _{\alpha}-\phi _{\alpha}^{\prime }\right) -\frac{3}{2}
\phi _{\alpha}\right) \mathbf{u}$. The equations (\ref{EqfiKT}) are solved subject to the constraint $\int f_{A}^Km_{A}\mathbf{u}d\mathbf{u+}\int f_{B}^Km_{B}\mathbf{u}d\mathbf{u}=0$ using the same computer aided method applied to the solution of the HCS. The functions $%
\Phi _{\alpha }^{K,T}$ are represented by series of Sonine
polynomials $\Phi _{\alpha }^{K,T}=\sum_{p=0}\frac{\widehat{k}_{\alpha
}^{K,T,p}}{n\sigma _{AB}^{2}}S_{\frac{3}{2}}^{p}\left( \gamma _{\alpha
}u^{2}\right) $ (the factor $\nu <1$ in the function $f_{\alpha }^{M,\nu }$
ensuring the convergence of the series), which is substituted in (\ref{EqfiKT}).
% to yield:
%\begin{eqnarray}
%&\sum_{q}&\bigg[\left( L_{\alpha \alpha }^{(1)}+L_{\alpha \alpha }^{(2)}+L_{\alpha \beta
%}^{(1)}\right) \left( S_{\frac{3}{2}}^{q}\left( \gamma _{\alpha }u^{2}\right) \sqrt{\gamma _{\alpha }%}\mathbf{u}%
%\right) +L_{\alpha \beta }^{(2)}\left( S_{\frac{3}{2}}^{q}\left( \gamma _{\alpha }u^{2}\right)\sqrt{\%gamma
%_{\beta }}\mathbf{u}\right) \nonumber\\
%&-&\frac{\sqrt{6\pi }}{9}\frac{n\sigma _{\alpha
%\beta }^{2}}{\sqrt{m_{0}}}\widetilde{\Gamma }\sqrt{T}\mathbf{H}_{\alpha
%}^{T}\left(S_{\frac{3}{2}}^{q}\left( \gamma _{\alpha }u^{2}\right)\sqrt{\gamma _{\alpha }}\mathbf{u}\%right)\bigg] =%
%\mathbf{R}_{\alpha }^{T}
%\end{eqnarray}
 The resulting equation is then projected on  $S_{\frac{3}{2}}^{N}\left( \gamma _{\alpha }u^{2}\right) 
\sqrt{\gamma _{\alpha }}\mathbf{u}$ to yield a linear system of equations for the
coefficients $\widehat{k}_{\alpha }^{K,T,p}$.  The system of equations
is then solved (after evaluating the coupling constants using the 
generating function method explained above). The resulting perturbations
are then used to calculate the transport coefficients. Results 
for the thermal diffusion constant, $\kappa_A^{T}$, and the
heat conductivity, $\lambda^{T}$, are  presented in Fig.~\ref{TR_fig}(b,c).
Note the nontrivial dependence of these transport  coefficients on
$\frac{m_A}{m_A+m_B}$. Although clearly equipartition is strongly broken,  these
transport coefficients are well behaved (as are those not shown here).
\section{Elastic energy and ``heat flux''}

\label{sec:elasticity_derivation} 

Consider a set of particles whose binary interaction potential is harmonic (or
approximated by): $
U\left[\vec{r}_{ij}(t)\right]=\frac{1}{2}K_{ij}\left(|\vec{r}_{ij}|-l_{ij}\right)^{2}$,
where $l_{ij}$ is the equilibrium separation of particles $i$ and $j$ and ${\bf
  r}_{ij} \equiv {\bf r}_i-{\bf r}_j$. The force on particle $i$ exerted by
particle $j$ is given by $ \vec{f}_{i/j} =
-\mbox{\boldmath$\nabla$}_{i}U\left(\vec{r}_{ij}\right) =
-K_{ij}\left(|\vec{r}_{ij}|-l_{ij}\right)\hat{\vec{r}}_{ij}$, where
$\hat{\vec{r}}_{ij} \equiv \vec{r}_{ij} / \left| \vec{r}_{ij} \right|$ is a
unit vector. Consider, for simplicity, an unstressed reference configuration,
in which all particle pairs are at their equilibrium separation
($\left|\vec{r}_{ij}^{0}\right|=l_{ij}$), where the superscript~$0$ denotes the
reference configuration. The relative displacement of particles $i$ and $j$ is
defined by: ${\bf u}_{ij} \equiv {\bf r}_{ij}-{\bf r}_{ij}^0$.  To linear order
in the displacements the corresponding force can be approximated by:
\begin{eqnarray}
  \vec{f}_{i/j} & = & -K_{ij}\left(|\vec{r}_{ij}|-l_{ij}\right)\hat{\vec{r}}_{ij}=-K_{ij}\left(|\vec{r}_{ij}^{0}+\vec{u}_{ij}|-\left|\vec{r}_{ij}^{0}\right|\right)\hat{\vec{r}}_{ij}\nonumber \\
   & = &
  -K_{ij}\left(\hat{\vec{r}}_{ij}^{0}\cdot\vec{u}_{ij}\right)\hat{\vec{r}}_{ij}^{0}.\label{eq:harmonic_force_linear}\end{eqnarray}

For atomic systems, the harmonic approximation is obtained from a Taylor
expansion of the (effective) interatomic potential around its minimum. For
granular materials, the contact interactions among the particles are typically
described using contact mechanics~\cite{Gladwell80,Johnson85} (force models
used in simulations of granular materials are reviewed
in~\cite{Sadd93,Walton95,Schafer96,Wolf96,Herrmann98}). For the present
discussion, it is assumed that the interparticle forces can be linearized for
small deformations with respect to a reference
configuration.

Consider the contact stress, see Eq.~(\ref{contactdef}). Upon substituting
the formula for the force to linear order in the displacements, Eq.~(\ref{eq:harmonic_force_linear}), 
in Eq.~(\ref{contactdef}) one obtains an expression for the linear elastic
stress tensor: 
\begin{eqnarray} \sigma_{\alpha\beta}^{\mathrm{lin}}(\vec{r},t) & = &
  \frac{1}{2}\sum_{ij}K_{ij}\hat{r}_{ij\alpha}^{0}r_{ij\beta}^{0}\hat{r}_{ij\gamma}^{0}u_{ij\gamma}\int_{0}^{1}ds\,\phi[\vec{r}-\vec{r}_{i}^{0}+s\vec{r}_{ij}^{0}].\label{eq:lin_stress}
\end{eqnarray}
It is interesting to examine the potential energy density as read off
Eq.~(\ref{e1}) and compare it to the relation used in continuum linear
elasticity.  In the quasistatic limit, the energy density reduces to the
potential energy density {[}in this limit, the time of deformation goes to
infinity ($t_f\to\infty$) and the velocities go to zero ($\vec{v}_{i}\to0$), so
that the kinetic energy is zero while the displacement
$\vec{u}_{i}\sim\vec{v}_{i} t_f$, is kept finite{]}. To lowest nonvanishing
order in the strain, the potential energy is given by
$\frac{1}{2}K_{ij}\left(\hat{\vec{r}}_{ij}^{0}\cdot\vec{u}_{ij}\right)^{2}$.
Hence, at this order:
\begin{equation}
  e^{\mathrm{el}}(\vec{r},t)=\frac{1}{4}\sum_{i,j}K_{ij}\left(\hat{\vec{r}}_{ij}^{0}\cdot\vec{u}_{ij}\right)^{2}\phi[\vec{r}-\vec{r}_{i}(t)].\label{eq:lin_elasticenergydensity}\end{equation}
Classical linear elasticity identifies the elastic energy 
as $e^{\mathcal{\mathrm{el}}}=\frac{1}{2}\mbox{\boldmath$\sigma$}^{\mathrm{lin}}:\mbox{\boldmath$\epsilon$}^{\mathrm{lin}}$,  where $\epsilon^{\mathrm{lin}}$ is defined in 
Eq.~(\ref{epslin}). 
Since
the stress is symmetric for central forces (it is easily verified that 
Eq.~(\ref{eq:lin_stress}) is invariant to the exchange of  $\alpha$
and $\beta$), we can calculate
$\frac{1}{2}\sigma_{\alpha\beta}^{\mathrm{lin}}\frac{\partial
  u_{\alpha}^{\mathrm{lin}}}{\partial r_{\beta}}=\frac{1}{2}\frac{\partial}{\partial
  r_{\beta}}
\left(\sigma_{\alpha\beta}^{\mathrm{lin}}u_{\alpha}^{\mathrm{lin}}\right) $
instead, 
the second equality following from the static equilibrium relation 
$\frac{\partial \sigma_{\alpha\beta}^{\mathrm{lin}}}{\partial r_\beta}=0$,
see also  Eq.~(\ref{second}). It follows, using 
Eq.~(\ref{contactdef}) and definition   (\ref{linearu}) that: 
\begin{displaymath}
\frac{1}{2} \sigma_{\alpha\beta}^{\mathrm{lin}} \epsilon^{\mathrm{lin}}_{\alpha\beta}
  = -\frac{1}{4}\frac{\partial}{\partial r_{\beta}}\sum_{ij}u_{\alpha}^{\mathrm{lin}} f_{i/j\alpha} r_{ij\beta}^{0}\int_{0}^{1}ds\,\phi[\vec{r}-\vec{r}_{i}^{0}+s\vec{r}_{ij}^{0}]
\end{displaymath}
Next, using Eq.~(\ref{flu}) one obtains: 
%\end{document}
\begin{displaymath}
 \frac{1}{2} \sigma_{\alpha\beta}^{\mathrm{lin}} \epsilon^{\mathrm{lin}}_{\alpha\beta} = -\frac{1}{4}\frac{\partial}{\partial r_{\beta}}\sum_{ij}\left[u_{i\alpha}- u'_{i\alpha}\right]
f_{i/j\alpha} r_{ij\beta}^{0}\int_{0}^{1}ds\,\phi[\vec{r}-\vec{r}_{i}^{0}+s\vec{r}_{ij}^{0}]
\end{displaymath}
\begin{displaymath}
  =  -\frac{1}{4}\sum_{ij}u_{i\alpha} f_{i/j\alpha} r_{ij\beta}^{0}
\frac{\partial}{\partial r_{\beta}}\int_{0}^{1}ds\,\phi[\vec{r}-\vec{r}_{i}^{0}+s\vec{r}_{ij}^{0}]
\end{displaymath}
\begin{displaymath}
 +\frac{1}{4}\frac{\partial}{\partial r_{\beta}}\sum_{ij}u'_{i\alpha}f_{i/j\alpha}r_{ij\beta}^{0}\int_{0}^{1}ds\,\phi[\vec{r}-\vec{r}_{i}^{0}+s\vec{r}_{ij}^{0}]
\end{displaymath}
\begin{displaymath}
  = \frac{1}{4}
\sum_{ij}u_{i\alpha}f_{ij\alpha}
\left(\phi\left[ {\bf r}-{\bf r}_i(t)  \right]
-\phi \left[  {\bf r}-{\bf r}_j(t)     \right]
\right)
\end{displaymath}
\begin{displaymath}
  +\frac{1}{4}\frac{\partial}{\partial r_{\beta}}\sum_{ij}u'_{i\alpha}f_{i/j\alpha}r_{ij\beta}^{0}\int_{0}^{1}ds\,\phi[\vec{r}-\vec{r}_{i}^{0}+s\vec{r}_{ij}^{0}]
\end{displaymath}
In equilibrium $\sum_j f_{i/j\alpha}=0$, hence 
$\sum_j f_{i/j\alpha}\phi\left( {\bf r}-{\bf r}_i(t)  \right)=0$ and it can be
replaced by $-\sum_j f_{i/j\alpha}\phi\left( {\bf r}-{\bf r}_i(t)  \right)$. 
Therefore 
\begin{displaymath}
\frac{1}{2} \sigma_{\alpha\beta}^{\mathrm{lin}} \epsilon^{\mathrm{lin}}_{\alpha\beta}=
-\frac{1}{4}
\sum_{ij}u_{i\alpha}f_{ij\alpha}
\left(\phi\left[ {\bf r}-{\bf r}_i(t)  \right]
+\phi \left[  {\bf r}-{\bf r}_j(t)     \right]
\right)
\end{displaymath}
\begin{displaymath}
  +\frac{1}{4}\frac{\partial}{\partial r_{\beta}}\sum_{ij}u'_{i\alpha}f_{i/j\alpha}r_{ij\beta}^{0}\int_{0}^{1}ds\,\phi[\vec{r}-\vec{r}_{i}^{0}+s\vec{r}_{ij}^{0}]
\end{displaymath}
\begin{displaymath}
  =  -\frac{1}{4}\sum_{ij}u_{ij\alpha}f_{i/j\alpha}\phi \left[  {\bf r}-{\bf r}_i(t)    \right]
  +\frac{1}{4}\frac{\partial}{\partial r_{\beta}}\sum_{ij}u'_{i\alpha}f_{i/j\alpha}r_{ij\beta}^{0}\int_{0}^{1}ds\,\phi[\vec{r}-\vec{r}_{i}^{0}+s\vec{r}_{ij}^{0}]
\end{displaymath}
Next, substituting the expression for the force, Eq.~(\ref{eq:harmonic_force_linear}), one obtains:
\begin{displaymath}
\frac{1}{2} \sigma_{\alpha\beta}^{\mathrm{lin}} \epsilon^{\mathrm{lin}}_{\alpha\beta} =  \frac{1}{4}\sum_{ij}K_{ij}\left(\hat{\vec{r}}_{ij}^{0}\cdot\vec{u}_{ij}\right)^{2}
\phi \left[ {\bf r}-{\bf r}_i(t) \right]
\end{displaymath}
\begin{displaymath}
+\frac{1}{4}\frac{\partial}{\partial r_{\beta}}\sum_{ij}f_{i/j\alpha}u'_{i\alpha}r_{ij\beta}^{0}\int_{0}^{1}ds\,\phi[\vec{r}-\vec{r}_{i}^{0}+s\vec{r}_{ij}^{0}]
\end{displaymath}
\begin{displaymath}
  =  e^{\mathrm{{el}}}(\vec{r},t)+\frac{1}{4}\frac{\partial}{\partial r_{\beta}}\sum_{ij}f_{i/j\alpha}u'_{i\alpha}r_{ij\beta}^{0}\int_{0}^{1}ds\,\phi[\vec{r}-\vec{r}_{i}^{0}+s\vec{r}_{ij}^{0}]
\end{displaymath}
\begin{equation}
  =  e^{\mathrm{{el}}}(\vec{r},t)+\frac{1}{8}\frac{\partial}{\partial r_{\beta}}\sum_{ij}f_{i/j\alpha}\left( u'_{i\alpha}
+  u'_{j\alpha}\right) r_{ij\beta}^{0}
\int_{0}^{1}ds\,\phi[\vec{r}-\vec{r}_{i}^{0}+s\vec{r}_{ij}^{0}] \label{eee}
\end{equation}
This result is identical to that obtained on the basis of dynamical considerations in Appendix B, cf. Eqs.~(\ref{tiny},\ref{last}).

It follows that the coarse grained elastic energy density is not precisely
given by the classical expression
$\frac{1}{2}\mbox{\boldmath$\sigma$}^{\mathrm{lin}}:\mbox{\boldmath$\epsilon$}^{\mathrm{lin}}$; the above
additional term provides a correction to the classical expression. The
correction represents the adiabatic limit of the divergence of the (time
integral of) the heat flux or, in physical terms, the divergence of the
fluctuating part of the work of the interparticle forces, i.e. the work done on
the fluctuating parts of the displacements or the work which is unresolved by
macroscopic elasticity. As this term is a divergence of a flux, its average
over a volume, $\Omega$, whose linear dimension is $W$, is proportional to
$W^{D-1}$, while the average over the first term is proportional to $W^{D}$.
Therefore their ratio tends to zero as $1/W$, i.e., the \emph{average} of
$\frac{1}{2}\mbox{\boldmath$\sigma$}^{\mathrm{lin}}:\mbox{\boldmath$\epsilon$}^{\mathrm{lin}}$ over a sufficiently
large volume (not its `local value') begets the standard elastic energy
density.  It should be emphasized that both contributions to the elastic energy
density are of the same (second) order in the displacements (or strain), i.e.
the additional term is not a ``higher order'' correction but rather a
correction that appears when the deformation is not affine (i.e., when ${\bf
  u}^\prime_{i} \ne 0$).
\subsection{Numerical results: 2D disordered harmonic networks}

In order to examine the correction to the classical elastic energy numerically,
we use disordered harmonic networks in 2D, obtained as follows: ``particles''
(of equal mass) are placed at the nodes of a triangular lattice, of lattice
constant $d$. The system is square shaped.  Nearest neighbors are connected by
springs (with rest lengths equal to the lattice constant, ensuring a
stress-free reference state). We introduce disorder in two ways: first, the
spring network is diluted by randomly removing a fraction $c$ of the springs.
Second, we employ spring constants which are uniformly distributed in the range
$[K-\delta K,K+\delta K]$. The particle positions are the same as in the
triangular lattice, hence the density remains uniform. We use periodic boundary
conditions in both axes, and apply a specified global strain to the system by
imposing a change in the period as well as (for shear deformation) Lees-Edwards
boundary conditions.

The (linear) static equilibrium equations are solved (by inversion of the
dynamical matrix) for a given applied global strain,
$\epsilon^{\mathrm{app}}_{\alpha\beta}$, yielding a set of corresponding
displacements $\left\{ \vec{u}_{i}\right\} $.  The latter are used for
calculating the CG linear strain field, $\epsilon^{\mathrm{lin}}$ [using
Eq.~(\ref{epslin})], the CG linear stress, $\sigma^{\mathrm{lin}}$
[Eq.~(\ref{eq:lin_stress}], the CG potential energy density, $e^{\mathrm{el}}$
[Eq.~(\ref{eq:lin_elasticenergydensity})], and the fluctuating energy flux,
$\frac{1}{2}\vec{Q}^{\mathrm{{force,rev,I}}}$ [Eq.~(\ref{last})].  The coarse
graining function used~\cite{Goldhirsch02} is a 2D Gaussian,
$\phi(\vec{r})=\frac{1}{\pi w^{2}}e^{-(|\vec{r}|/w)^{2}}$.

We verified that for a lattice with uniform spring constants (i.e., $c=\delta
K=0$), the strain components are uniform and equal to the applied global strain
for $w\gtrsim d$, as expected, since in for this system, under homogeneous
deformation, the particle displacements are affine. The resulting
stress components are uniform as well, and are consistent with the continuum
isotropic elastic moduli for a triangular lattice (Lam\'{e} constants
$\lambda=\mu=\frac{\sqrt{3}K}{4}$). As expected, in this case the CG energy
density conforms to the classical expression,
$e^{\mathrm{el}}=\frac{1}{2}\mbox{\boldmath$\sigma$}^{\mathrm{lin}}:\mbox{\boldmath$\epsilon$}^{\mathrm{lin}}$.

\begin{figure}
\begin{center}\begin{tabular}{cc}
\subfigure[]{\includegraphics[%
  height=0.42\hsize,clip]{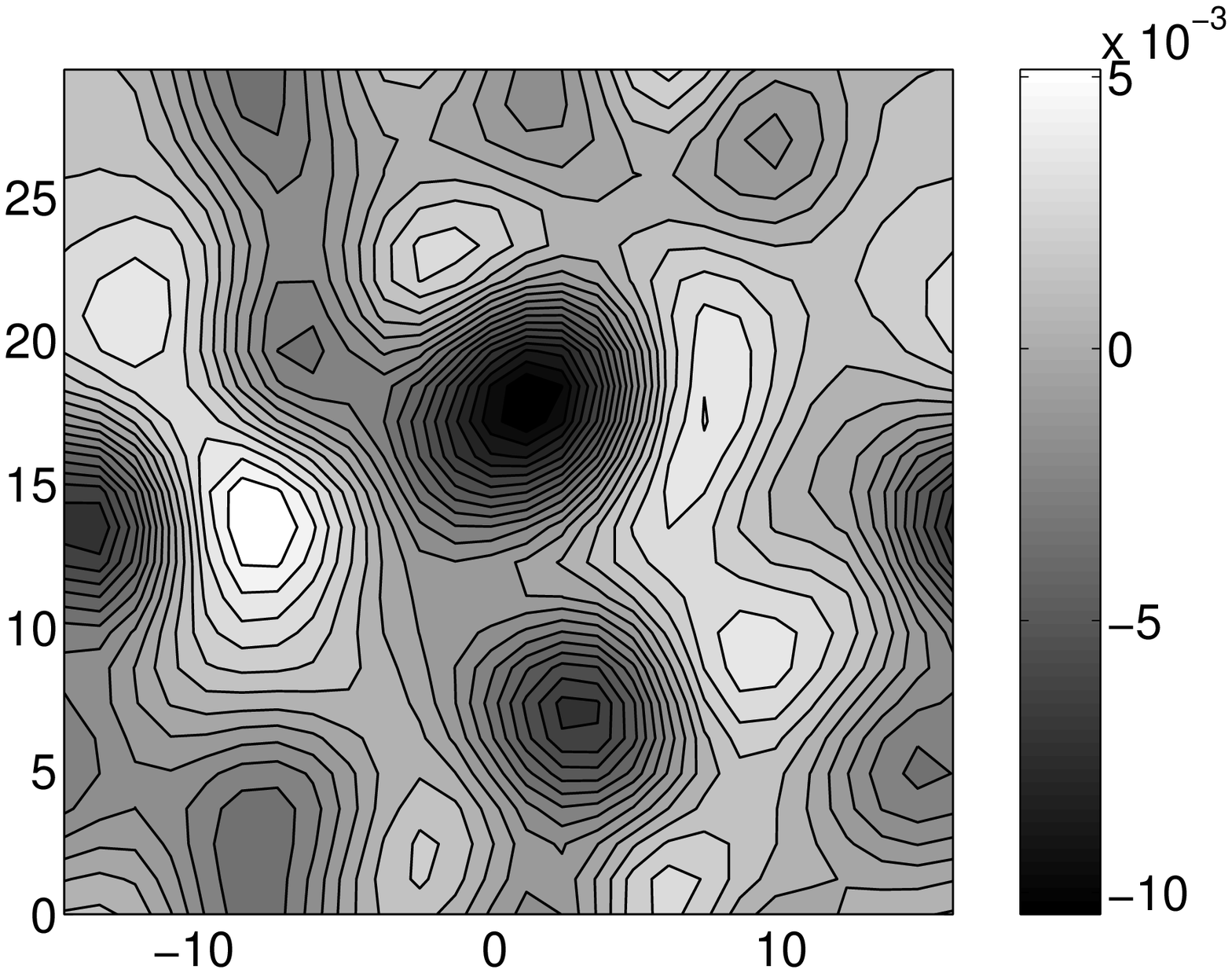}\label{fig:dErel}}&
\subfigure[]{\includegraphics[%
  height=0.39\hsize,clip]{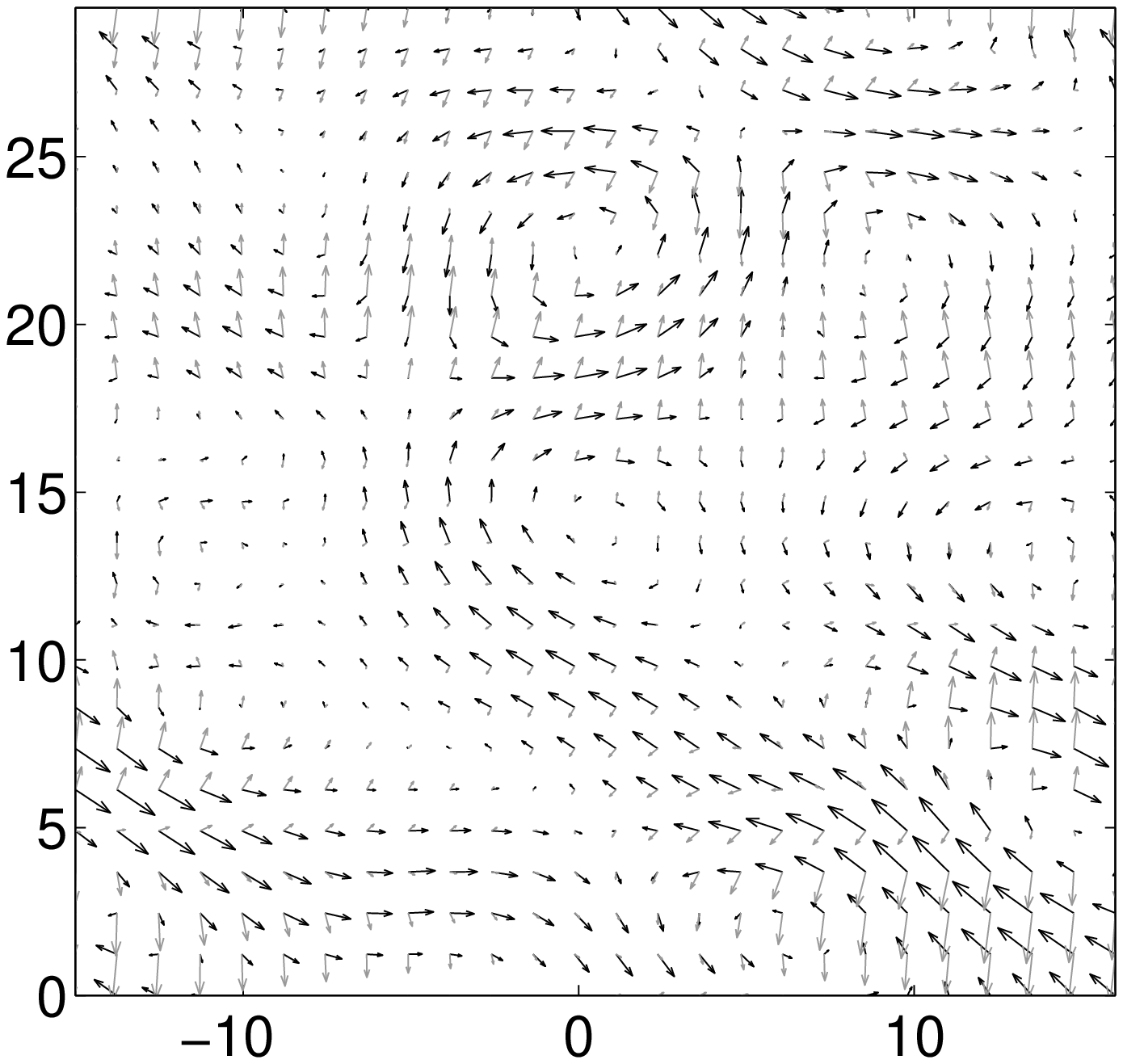}\label{fig:Q_gradN}}\tabularnewline
\end{tabular}\end{center}
\caption{(a) A contour plot of the relative difference between the elastic
  energy density, $e^{\mathrm{el}}$, and its ``classical'' value,
  $\frac{1}{2}\mbox{\boldmath$\sigma$}^{\mathrm{lin}}:\mbox{\boldmath$\epsilon$}^{\mathrm{lin}}$, in a disordered
  harmonic network (see text). (b) The fluctuating energy flux,
  $\frac{1}{2}\vec{Q}^{\mathrm{{force,rev,I}}}$ (black) and the (negative)
  gradient of the ``noise'', $-\mbox{\boldmath$\nabla$} \eta$ (gray), in the same system (arbitrary
  scaling).}
\label{fig:dErel_Q_gradN}
\end{figure}

For disordered systems, even under a ``homogeneous applied strain'' as
described above, there is a correction to the classical expression for the
energy density, as detailed above. Fig.~\ref{fig:dErel} presents a contour plot
of the relative energy difference $\Delta \tilde{e}\equiv 2
\frac{e^{\mathrm{el}}}{\mbox{\boldmath$\sigma$}^{\mathrm{lin}}:\mbox{\boldmath$\epsilon$}^{\mathrm{lin}}}-1=-
\frac{\mathrm{div}\, {\bf
    Q}^{\mathrm{{force,rev,I}}}}{\mbox{\boldmath$\sigma$}^{\mathrm{lin}}:\mbox{\boldmath$\epsilon$}^{\mathrm{lin}}}$,
for a system of $30\times 30$ particles, with $c=2\%$ and $\delta K/K=25\%$
subject to an applied strain $\epsilon^{\mathrm{app}}_{xx}=-\epsilon^{\mathrm
  app}_{yy}=5\cdot10^{-3}, \epsilon^{\mathrm{app}}_{xy}=0$, calculated with a
CG width, $w=4d$. The deviation from the classical expression is non-negligible
($\Delta \tilde{e}\simeq 1\%$), even thought the present system has a relatively
small disorder [we verified that in the same system, a linear stress-strain
relation holds to within $4\%$~\cite{Goldhirsch02}]. It is interesting to try
to relate the fluctuating energy flux to a local measure of the disorder in the
particle displacements. A possible field for characterizing the fluctuating
displacements is the ``noise'' field~\cite{Goldenberg06c}, $\eta({\bf r})\equiv
\sum_i m_i |\vec{u}'_i(\vec{r})|^2 \phi[\vec{r}-\vec{r}_i(t)]$, defined in an
analogous way to the kinetic temperature. In Fig.~\ref{fig:Q_gradN} we compare
the fluctuating energy flux with the gradient of the noise field. It appears
that their magnitude (but not their directions) exhibits quite a strong
correlation, suggesting that the fluctuating energy flux may indeed be related
to gradients of the noise, but the corresponding ``conductivity'' may be
anisotropic. This is to be expected, since the system is locally anisotropic
[as the local elastic tensor has been checked to be
anisotropic~\cite{Goldhirsch02}]. It is possible that a tensorial, rather than
scalar, characterization of the displacement fluctuations is required,
analogous to an ``anisotropic temperature''.
\section{Conclusion}
We have shown that the classical definition of granular temperature as a
measure of the fluctuating kinetic energy is useful for the description of
granular gases even when this temperature is not the (inverse of the)
thermodynamic conjugate potential of the energy, as in (local) equilibrium.  An
analysis of the case of a near-elastic granular gas shows that indeed one can
venture into the nonequilibrium domain starting from local equilibrium (for the
near elastic case). The example of a binary granular gas mixture in which the
temperatures of the components can be very different from each other and still
the Chapman-Enskog method produces sensible results, is a compelling case in
favor of the use of the granular temperature field for far-from-equilibrium
systems, at least in the fluidized phase.  In the `opposite' limit of a
granular solid we have shown that the heat flux (which is always present, even
in athermal system, as shown in Appendix A) is not `lost' when one performs a
deformation of an elastic solid: its effect appears in the form of a correction
to the classical elastic energy density. This correction is strongly correlated
with the gradient of an `elastic temperature' which is a measure of the degree
of non-affinity induced by the randomness of the system. Note that the system
considered is not pre-stressed whereas many granular systems are. This fact
does not change the fundamental conclusion that in elastic solids there is an
accumulation of energy as a result of work done on the non-affine (or
fluctuating) degrees of freedom. As the physics of this finding is very general
we believe it applies (with a proper modification) to granular solids beyond
the elastic regime. Indeed a strong correlation of the displacement `noise'
with the appearance of localized plastic events has been suggested in
\cite{Goldenberg06c}. Another connection of this kind has been proposed in
\cite{Lemaitre02}. At this stage we do not know whether there is a direct link
between this `new' contribution that stems from the heat flux and (gradients
of) `configurational temperatures'. However, since there is an obvious link
between diffusion and non-affinity (see e.g., \cite{Utter07}) and a seemingly
established link between diffusion and a `configurational temperature' (through
an FD relation) it comes to reason that the `extra' term in the elastic energy
is related to a configurational measure of disorder. Following the spirit of a
suggestion by Savage \cite{Savage98} one may perhaps conjecture that an
effective temperature which combines the configurational and kinetic
contributions (perhaps just as a sum) can serve as a generalized granular
temperature valid for all granular phases (with appropriate extensions to other
systems). At this stage it seems that much more research is necessary to
establish whether this or similar conjectures are of any physical value.  As a
final comment we would like to mention the rather well known fact that (even
relatively simple) non-equilibrium systems exhibit rich behavior and cannot
usually be characterized by the same number of macroscopic variables or fields
as (near) equilibrium systems. Therefore, even if it turns out that many
properties of granular matter can be correlated with a generalized temperature
or temperatures there is a good chance that more characterizations will be
needed for this class of systems to fully describe or explain their behavior.

\appendix
\section{Continuum Mechanics: a microscopic derivation}

The equations of continuum mechanics are derived below from 
microscopic considerations. 
For sake of simplicity it is assumed that the constituents experience
binary interactions whose reversible part is determined by a potential. The
derivation below is a slight extension of what can be found
in \cite{Glasser01,Babic97} but unlike in \cite{Glasser01} we do not invoke 
temporal coarse graining, for simplicity. 

Consider a set of particles of mass $m_i$, whose (center of mass)
positions and velocities are
${\bf r}_i(t)$ and  ${\bf v}_i(t) \equiv \dot{\bf r}_i(t)$, 
respectively. Let the binary interaction potential be denoted by
$U({\bf r}_{ij})$, where ${\bf r}_{ij} \equiv {\bf r}_i - {\bf r}_j$. It
is assumed that $U=0$ for $i=j$.  In
addition, let $\phi({\bf r})$ denote a spatial coarse graining function, i.e. 
$\phi$ is normalized (its integral over space equals  unity), it is
positive semidefinite and has a single maximum at ${\bf r}=0$. It is often
convenient to employ a Gaussian  distribution for $\phi$. The width
of $\phi$ is the ``coarse graining scale''.

The coarse-grained (CG) mass density, $\rho({\bf r},t)$, at position ${\bf r}$
and time $t$ is given by: $ \rho({\bf r},t) \equiv \sum_i m_i \phi\left[{\bf
    r}-{\bf r}_i(t)\right] $.  Note that while ${\bf r}_i(t)$ is the position
of particle $i$ at time $t$, ${\bf r}$ is merely a considered position in
space. Similarly, define the CG momentum density, ${\bf p}({\bf r},t)$, by: $
{\bf p}({\bf r},t) \equiv \sum_i m_i {\bf v}_i (t) \phi\left[{\bf r}-{\bf
    r}_i(t)\right] $.  The CG energy density, $e({\bf r},t)$ is given by:
\begin{equation}
e({\bf r},t) \equiv \sum_i \frac{1}{2} m_i v^2_i \phi\left[ {\bf r}-{\bf r}_i(t)\right]+ \frac{1}{2} \sum_{i,j} U({\bf r}_{ij}(t))\phi\left[ {\bf r}-{\bf r}_i(t)\right] \label{e1}
\end{equation}
By taking the time derivative of the density one obtains: 
\begin{equation}
\dot{\rho}({\bf r},t)= -\frac{\partial}{\partial r_\alpha}
\sum_i m_i v_{i\alpha} (t) \phi\left[ {\bf r}-{\bf r}_i(t)\right]
= -\mathrm{div}\, {\bf p}({\bf r},t)\label{rhodot1}
\end{equation}
where the summation convention has been invoked and Greek letters denote
Cartesian coordinates. Defining the velocity field, ${\bf V}({\bf r},t)$,
as: $
%\begin{equation}
{\bf V}({\bf r},t) \equiv \frac{{\bf p}({\bf r},t)}{\rho({\bf r},t)} \label{veldef}$,
%\end{equation}
one obtains from Eq.~(\ref{rhodot1}) the equation of continuity:
$\dot{\rho} =-\mathrm{div}\, (\rho {\bf V})$. 
Next, upon taking the derivative of the momentum field one obtains: 
\begin{equation}
\dot{ p}_\alpha ({\bf r},t) = -\frac{\partial}{\partial r_\beta}
\sum_i m_i v_{i\alpha}(t) v_{i\beta} (t) \phi\left[ {\bf r}-{\bf r}_i(t)\right]
+ \sum_i m_i \dot{v}_{i\alpha} \phi\left[ {\bf r}-{\bf r}_i(t)\right]
\label{p1}
\end{equation}
Denote by ${\bf f}_i(t)$ the force experienced by particle $i$ and by
${\bf f}_{i/j}(t)$ the force exerted by particle $j$ on $i$.
Using Newton's second and third  laws one obtains: 
%\begin{displaymath}
$
\sum_i m_i \dot{ v}_{i\alpha}(t) \phi\left[ {\bf r}-{\bf r}_i(t)\right]
=\sum_{i,j} f_{i/j\alpha}(t) \phi\left[ {\bf r}-{\bf r}_i(t)\right]
%\end{displaymath}
%\begin{displaymath}
=\sum_{i,j} f_{j/i\alpha}(t) \phi\left[ {\bf r}-{\bf r}_j(t)\right]
= -\sum_{i,j} f_{i/j\alpha}(t) \phi\left[ {\bf r}-{\bf r}_j(t)\right]
%\end{displaymath}
$.
It follows that: 
\begin{displaymath}
\sum_i m_i \dot{ v}_{i\alpha}(t) \phi\left[ {\bf r}-{\bf r}_i(t)\right]
= 
\end{displaymath}
\begin{displaymath}
=\frac{1}{2}\sum_{ij} f_{i/j\alpha}(t) \left( \phi\left[ {\bf r}-{\bf r}_i(t)\right]
-\phi\left[ {\bf r}-{\bf r}_j(t)\right]
\right)
\end{displaymath}
\begin{displaymath}
= -\frac{1}{2} \sum_{ij} f_{i/j\alpha} (t) \int_0^1 ds \frac{\partial}{\partial s}
\phi\left[ {\bf r}- {\bf r}_i (t)  + s{\bf r}_{ij}(t) \right]
\end{displaymath}
\begin{equation}
=-\frac{\partial}{\partial r_\beta} 
\frac{1}{2} \sum_i m_i f_{i/j\alpha} r_{ij\beta}(t) \int_0^1
ds  \phi\left[ {\bf r}-{\bf r}_i(t)+s {\bf r}_{ij}(t)\right]
\end{equation}
Substituting the latter result in Eq.~(\ref{p1}) one obtains: 
%\begin{equation}
$
\dot{p}_\alpha({\bf r},t) = -\frac{\partial}{\partial r_\beta} P_{\alpha\beta}
({\bf r},t) \label{p2} 
%\end{equation}
$,
where the pressure tensor, $\mathbf P$, is given by:
\begin{displaymath}
P_{\alpha\beta}({\bf r},t) =
\sum_i m_i v_{i\alpha}(t) v_{i\beta}(t) 
\phi\left[ {\bf r}-{\bf r}_i(t)\right]
\end{displaymath}
\begin{equation}
+ \frac{1}{2}\sum_{ij} f_{i/j\alpha}({\bf r},t) r_{ij\beta}({\bf r},t)
\int_0^1 ds \phi\left[ {\bf r}-{\bf r}_i(t)+ s{\bf r}_{ij}(t) \right] \label{prev}
\end{equation}
Next, define the fluctuating part of the velocity of a particle (when
coarse graining is around the position ${\bf r}$ at time $t$) as:
${\bf v}^\prime_{i}({\bf r},t) \equiv {\bf v}_i(t) - {\bf V}({\bf r},t)$.
Substituting this definition in Eq.~(\ref{prev}) and performing straightforward algebra (using $\sum_i v^\prime_i({\bf r},t) \phi\left(
{\bf r}-{\bf r}_i(t)\right) =0 $, following the definitions of 
${\bf v}^\prime$ and the velocity field ${\bf V}$), one obtains:
\begin{displaymath}
P_{\alpha\beta}({\bf r},t) =
\rho({\bf r},t)  V_\alpha({\bf r},t)  V_\beta ({\bf r},t) 
+\sum_{i} m_i  v^\prime_{i\alpha} ({\bf r},t)  v^\prime_{i\beta} ({\bf r},t) 
\phi\left[ {\bf r}-{\bf r}_i(t)\right]
\end{displaymath}
\begin{equation}
+\frac{1}{2} \sum_{ij} f_{i/j\alpha}({\bf r},t) r_{ij\beta}({\bf r},t)
\int_0^1 ds \phi\left[ {\bf r}-{\bf r}_i(t)+ s {\bf r}_{ij} (t) \right]
\end{equation}
Identifying
\begin{equation}
\sigma^{\mathrm{{kin}}}_{\alpha\beta}({\bf r},t)  = -\sum_{i}m_i  v^\prime_{i\alpha} ({\bf r},t)  v^\prime_{i\beta} ({\bf r},t) 
\phi\left[ {\bf r}-{\bf r}_i(t)\right] \label{kineticstress1}
\end{equation}
 as the 
``kinetic stress'' and 
\begin{equation}
\sigma^{\mathrm{{cont}}}_{\alpha\beta}({\bf r},t) = 
 -\frac{1}{2} \sum_{ij} f_{i/j\alpha}({\bf r},t) r_{ij\beta}({\bf r},t)
\int_0^1 ds \phi\left[ {\bf r}-{\bf r}_i(t) +s {\bf r}_{ij} (t) \right]
\label{contactdef}
\end{equation}
 as the ``contact stress'', the stress
tensor, $\mbox{\boldmath $\sigma$}$,  is given by: 
\begin{displaymath}
\sigma_{\alpha\beta}({\bf r},t) =
-\sum_{i}m_i  v^\prime_{i\alpha} ({\bf r},t)  v^\prime_{i\beta} ({\bf r},t) 
\phi\left[ {\bf r}-{\bf r}_i(t) \right]
\end{displaymath}
\begin{equation}
-\frac{1}{2} \sum_{ij} f_{i/j\alpha}({\bf r},t) r_{ij\beta}({\bf r},t)
\int_0^1 ds \phi\left[ {\bf r}-{\bf r}_i(t)+s {\bf r}_{ij}(t)\right]
\label{stress}
\end{equation}
Note that an opposite convention for the sign of the stress tensor is
often employed in the field of granular matter. 
It follows from  Eq.~(\ref{stress}) (employing 
${\bf p} = \rho {\bf V}$ and invoking the equation of continuity)
that:
\begin{equation}
\rho \frac{DV_\alpha ({\bf r},t)}{Dt} = \frac{\partial \sigma_{\alpha\beta}
({\bf r},t) }{\partial r_\beta} \label{velvel}
\end{equation}
where $\frac{D{\bf V}}{Dt}\equiv \frac{\partial {\bf V}} {\partial
t} + {\bf V} \cdot \mbox{\boldmath$\nabla$} {\bf V}$ denotes the material derivative. 

Next, consider the energy density. Taking the derivative of 
Eq.~(\ref{e1}) one obtains after some simple algebra and the use of Newton's
second law: 
\begin{displaymath}
\dot{e}({\bf r},t) = -\frac{\partial}{\partial r_\alpha}
 \sum_i\bigg[  \frac{1}{2} m_i v_i^2 (t) 
 + \frac{1}{2} \sum_{j} U({\bf r}_{ij}(t)) \bigg] v_{i\alpha}(t)
\phi\left[{\bf r}-{\bf r}_i(t)\right]
\end{displaymath}
\begin{equation}
+ \sum_i {\bf f}_i(t) \cdot {\bf v}_i (t) \phi\left[{\bf r}-{\bf r}_i(t)\right]
-\frac{1}{2}\sum_i {\bf f}^{\mathrm{{rev}}}_{i/j}(t) \cdot {\bf v}_{ij}(t) 
\phi\left[{\bf r}-{\bf r}_i(t)\right]\label{e3}
\end{equation}
where the reversible part of the force ${\bf f}_{i/j}$ 
is given by ${\bf f}^{\mathrm{{rev}}}_{i/j}(t)=-\frac{\partial U({\bf r}_{ij}(t))}{\partial {\bf  r}_{ij}(t)}$. 
Using the above  decomposition of the velocity,
${\bf v}_i(t) = {\bf v}^\prime_i({\bf r},t) + {\bf V}({\bf r},t)$,  one
obtains that the  term whose divergence appears  in Eq.~(\ref{e3}) can be 
rewritten as: 
\begin{displaymath}
\sum_i \frac{1}{2} m_i v_i^2 (t) v_{i\alpha}(t) \phi\left[{\bf r}-{\bf r}_i(t)
\right] + \frac{1}{2} \sum_{ij} U({\bf r}_{ij}(t)) v_{i\alpha}(t)
\phi\left[{\bf r}-{\bf r}_i(t)
\right]
\end{displaymath}
\begin{displaymath}
= e({\bf r},t) {\bf V}_\alpha({\bf r},t) 
+ \frac{1}{2} \sum_i m_i v^{\prime 2}_{i}({\bf r},t) 
 v^{\prime}_{i\alpha}({\bf r},t) \phi\left[{\bf r}-{\bf r}_i(t)
\right]
\end{displaymath}
\begin{equation}
-V_\beta({\bf r},t) \sigma^{\mathrm{{kin}}}_{\alpha\beta}({\bf r},t)
+\frac{1}{2} \sum_{ij} U({\bf r}_{ij}(t))  v^{\prime}_{i\alpha}({\bf r},t) \phi\left[{\bf r}-{\bf r}_i(t)
\right] \label{term1}
\end{equation}
Next, note that
\begin{displaymath}
\sum_i {\bf f}_i(t) \cdot {\bf v}_i(t)\phi\left[{\bf r}-{\bf r}_i(t)\right]
=\sum_{ij} {\bf f}_{i/j} (t) \cdot {\bf v}_i(t)\phi\left[{\bf r}-{\bf r}_i(t)\right]
\end{displaymath}
\begin{displaymath}
= -\sum_{ij} {\bf f}_{i/j} (t) \cdot {\bf v}_j(t)\phi\left[{\bf r}-{\bf r}_j(t)\right]
\end{displaymath}
\begin{displaymath}
=\frac{1}{2} \sum_{ij} {\bf f}_{i/j} (t) \cdot {\bf v}_{ij}(t)
\phi\left[{\bf r}-{\bf r}_i(t)\right]
\end{displaymath}
\begin{displaymath}
+\frac{1}{4} \sum_{ij} {\bf f}_{i/j} (t) \cdot \left({\bf v}_i(t)+ {\bf v}_j(t) 
\right)
\left( \phi\left[{\bf r}-{\bf r}_i(t)\right] 
-\phi\left[{\bf r}-{\bf r}_j(t)\right] \right)
\end{displaymath}
\begin{displaymath}
=\frac{1}{2} \sum_{ij} {\bf f}_{i/j} (t) \cdot {\bf v}_{ij}(t)
\phi\left[{\bf r}-{\bf r}_i(t)\right]
\end{displaymath}
\begin{equation}
-\frac{\partial}{\partial r_\alpha}
\frac{1}{4} \sum_{ij} {\bf f}_{i/j} (t) \cdot \left[{\bf v}_i(t)+ {\bf v}_j(t) 
\right] {\bf r}_{ij\alpha} (t) 
\int_0^1 ds \phi\left[{\bf r}-{\bf r}_i(t)+ s{\bf r}_{ij} (t) \right] 
\label{term2}
\end{equation}
Using  velocity decomposition  and Eq.~(\ref{contactdef}) one  obtains from
Eq.~({\ref{term2}}):
\begin{displaymath}
\sum_i {\bf f}_i(t) \cdot {\bf v}_i(t)\phi\left[{\bf r}-{\bf r}_i(t)\right] = 
\frac{1}{2} \sum_{ij}{\bf f}_{i/j} (t) \cdot {\bf v}_{ij} (t) 
\phi\left[{\bf r}-{\bf r}_j(t)\right]
\end{displaymath}
\begin{displaymath}
-\frac{1}{4} \frac{\partial}{\partial r_\alpha}
\left(
\sum_{ij} {\bf f}_{i/j}(t)\cdot \left[ {\bf v}^\prime_i({\bf r},t)
+ {\bf v}^\prime_j({\bf r},t) \right] r_{ij\alpha}(t)
\int_0^1 ds \phi\left[{\bf r}-{\bf r}_i(t) + s {\bf r}_{ij}(t)\right]
\right)
\end{displaymath} 
\begin{equation}
+ \frac{\partial}{\partial r_\alpha}\bigg( {\bf V}_\beta({\bf r},t)
\sigma^{\mathrm{{cont}}}_{\beta\alpha}({\bf r},t)\bigg) \label{term4}
\end{equation}
 Next, combining
Eqs. (\ref{stress},\ref{e3},\ref{term4}) one obtains: 
\begin{equation}
\dot{e}({\bf r},t) = - \frac{\partial} {\partial r_\alpha}
\bigg(
e({\bf r},t) {\bf V}_\alpha({\bf r},t) 
-{\bf V}_\beta({\bf r},t) \sigma_{\beta\alpha}({\bf r},t) + 
 Q_\alpha({\bf r},t) 
\bigg)- \rho({\bf r},t) \Gamma({\bf r,t}) \label{energy}
\end{equation}
where ${\bf Q}$ is identified as the heat flux and is given by:
\begin{displaymath}
Q_\alpha =
\frac{1}{2}  \sum_{ij} \left(
U({\bf r}_{ij}(t)) +  m_i v_i^{\prime 2} v^\prime_{i\alpha} \right)
v^\prime_{i\alpha}({\bf r},t) \phi\left[ {\bf r}-{\bf r}_i(t)\right]
\end{displaymath}
\begin{equation}
+\frac{1}{4} \sum_{ij} {\bf f}_{i/j}(t)\cdot \bigg( {\bf v}^\prime_i({\bf r},t)
+ {\bf v}^\prime_j({\bf r},t) \bigg) r_{ij\alpha}(t)
\int_0^1 ds \phi\left[{\bf r}-{\bf r}_i(t) + s {\bf r}_{ij}(t)\right]
\label{heatflux1}
\end{equation}
and the sink term,  $-\rho\Gamma$ is given by:
\begin{displaymath}
-\rho({\bf r},t) \Gamma({\bf r},t) = \frac{1}{2} \sum_{ij} \bigg( {\bf f}_{i/j}({\bf r},t) 
- {\bf f}^{\mathrm{{rev}}}_{ij}({\bf r},t) \bigg)\cdot {\bf v}_{ij} ({\bf r},t) \phi\left[ {\bf r}-{\bf r}_i(t)\right]
\end{displaymath}
\begin{equation}
= \frac{1}{2}  \sum_{ij}  {\bf f}^{\mathrm{{irrev}}}_{ij}({\bf r},t) 
\cdot {\bf v}_{ij} ({\bf r},t) \phi\left[ {\bf r}-{\bf r}_i(t)\right]
\end{equation}
where the dissipative part of the force, 
 ${\bf f}^{\mathrm{{irrev}}}_{ij}$, 
is given by ${\bf f}_{i/j}-{\bf f}^{\mathrm{{rev}}}_{ij}$. 
Note that the heat flux represents the transport of energy by the fluctuating
velocity and the power related to the fluctuating velocities. The
energy sink term is entirely due to the power associated with the
dissipative part of the forces. 

The energy density can be rewritten as follows:
\begin{displaymath}
e({\bf r},t) \equiv \sum_i \frac{1}{2} m_i v^{\prime 2}_i \phi\left[ {\bf r}-{\bf r}_i(t)\right]
\end{displaymath}
\begin{equation}
+\frac{1}{2} \rho({\bf r}) {\bf V}^2 ({\bf r},t)
 +\frac{1}{2} \sum_{i,j} U({\bf r}_{ij}(t))\phi\left[ {\bf r}-{\bf r}_i(t)\right] \label{e10}
\end{equation}
Denote  the internal specific energy by $\overline{u}$. The latter
is given by  
\begin{equation}
\rho({\bf r},t) \overline{u}({\bf r},t)  \equiv \sum_i \frac{1}{2} m_i 
v^{\prime 2}_i \phi\left[ {\bf r}-{\bf r}_i(t)\right] +
\frac{1}{2} \sum_{i,j} U({\bf r}_{ij}(t))\phi\left[ {\bf r}-{\bf r}_i(t)\right]
\label{defT}
\end{equation}
and the energy density therefore equals  $\frac{1}{2} \rho {\bf V}^2 + \rho \overline{u}$. Substituting this decomposition into Eq.~(\ref{energy}) and using Eqs.~(\ref{rhodot1},\ref{velvel}) one obtains: 
\begin{equation}
\rho({\bf r},t) \frac{D\overline{u}({\bf r},t)}{Dt}
= \frac{\partial V_\beta({\bf r},t)}{\partial r_\alpha} \sigma_{\beta\alpha}
({\bf r},t) - \mathrm{div}\, {\bf Q}({\bf r},t) -\rho({\bf r},t) \Gamma({\bf r},t) \label{internal}
\end{equation}
\section{The quasistatic limit of Continuum Mechanics}
Consider the case of  slow deformations of a  near-static 
assembly of particles. To this end, let the particle velocities be denoted
by ${\bf v}_i(t) = \delta {\bf \tilde v}_i(t)$, where $\delta$ is a small
parameter and  $\tilde{A}$ is considered to be $O(1)$ 
for any dynamical variable, $A$. The  velocity field, ${\bf V}$, as
well as all other fields are 
rescaled in a similar way. 
Furthermore, let time be expressed as $\tilde{t}= \frac{t}{\delta}$, hence
$\frac{\partial} {\partial t} = \delta \frac{\partial}{{\partial \tilde{ t}}}$. With this 
rescaling the equations of continuum mechanics, Eqs.~(\ref{rhodot1},\ref{velvel},\ref{energy})
become: 
\begin{displaymath}
\delta \frac{\partial \rho}{\partial \tilde{t}} =
 - \delta \mathrm{div}\,(\rho {\bf \tilde{V}})
\end{displaymath}
\begin{displaymath}
\delta^2 \rho \frac{D\tilde{V}_\alpha}{D\tilde{t}} =
\delta^2 \frac{\partial\tilde{\sigma}^{\mathrm{{kin}}}_{\alpha\beta}}{\partial r_\beta}
+ \frac{\partial \sigma^{\mathrm{{cont}}}_{\alpha\beta}}{\partial r_\beta}
\end{displaymath}
\begin{displaymath}
\delta^3 \frac{\partial \tilde{e}^{\mathrm{{kin}}}}{\partial \tilde{t}}
+\delta \frac{\partial \tilde{e}^{\mathrm{{pot}}}}{\partial \tilde{t}}
=- \frac{\partial}{\partial r_\alpha}
\bigg(
\delta^3 \tilde{e}^{\mathrm{{kin}}}{\bf \tilde{V}}_\alpha 
+ \delta  e^{\mathrm{{pot}}} {\bf \tilde{V}}_\alpha 
-\delta^3 {\bf \tilde{V}}_\beta \tilde{\sigma}^{\mathrm{{kin}}}_{\beta\alpha}
\end{displaymath}
\begin{equation}
-\delta  {\bf \tilde{V}}_\beta {\sigma}^{\mathrm{{cont}}}_{\beta\alpha}
+ \delta \tilde{Q}^{\mathrm{{pot}}}_\alpha
+\delta^4 \tilde{Q}^{\mathrm{{kin}}}_\alpha
+\delta \tilde{Q}^{\mathrm{{force}}}_\alpha
\bigg)
- \delta \rho  \tilde{\Gamma}
\end{equation}
where the superscript `kin' refers to the kinetic part, the 
superscript `pot' refers to the potential part (involving the potential
$U$) and the superscript force refers to the fluctuating power contribution
to the heat flux. To linear order in $\delta$ the above equations reduce to: 
\begin{equation}
\frac{\partial \rho}{\partial \tilde{t}} =
 - \mathrm{div}\,(\rho {\bf \tilde{V}})\label{first}
\end{equation}
\begin{equation}
\frac{\partial \sigma^{\mathrm{{cont}}}_{\alpha\beta}}{\partial r_\beta} =0
\label{second} 
\end{equation}
\begin{equation}
 \frac{\partial \tilde{e}^{\mathrm{{pot}}}}{\partial \tilde{t}}
=- \frac{\partial}{\partial r_\alpha}
\bigg(
 e^{\mathrm{{pot}}} {\bf \tilde{V}}_\alpha 
-  {\bf \tilde{V}}_\beta {\sigma}^{\mathrm{{cont}}}_{\beta\alpha}
+ {Q}^{\mathrm{{pot}}}_\alpha
+ \tilde{Q}^{\mathrm{{force}}}_\alpha
\bigg)
-\rho \tilde {\Gamma}\label{slow11}
\end{equation}
Using Eq.~(\ref{second}), the equation for the energy
density becomes to linear order in the `slowness', $\delta$ (and 
after reverting to the original time variable):
\begin{equation}
 \frac{\partial {e}^{\mathrm{{pot}}}} {\partial {t}}
=- \frac{\partial}{\partial r_\alpha}
\bigg(
  e^{\mathrm{{pot}}} {\bf {V}}_\alpha 
+ {Q}^{\mathrm{{pot}}}_\alpha
+ {Q}^{\mathrm{{force}}}_\alpha
\bigg)
 + \frac{\partial {\bf {V}}_\beta}
{\partial r_\alpha}
  {\sigma}^{\mathrm{ {cont}}}_{\beta\alpha}
-\rho  {\Gamma}\label{slow1}
\end{equation}
We further specialize to the case when the particle relative
displacements are small and the potential is quadratic in the
relative displacements (to leading order). 
Let ${\bf u}_i(t)$ be the displacement of particle $i$: 
$\dot{\bf u}_i(t) = {\bf v}_i$. Also, define the displacement field
${\bf u}({\bf r},t)$ via the Lagrangian relation: 
$\frac{\partial {\bf u}^{\mathrm{{La}}}({\bf R},t)}{\partial t}
={\bf V}^{\mathrm{{La}}}({\bf R},t)$, where ${\bf R}$ is the initial position
of a pathline and the superscript `La' denotes the fact that one is employing
Lagrangian coordinates. As shown in \cite{Goldhirsch02}
\begin{equation}
{\bf u}^{\mathrm{{lin}}}({\bf r},t) =
\frac{\sum_i m_i {\bf u}_i(t) \phi\left[{\bf r}-{\bf r}_i(t)\right]}
{\sum_j m_j  \phi\left[{\bf r}-{\bf r}_j(t)\right]} \label{linearu}
\end{equation}
where the superscript `lin' denotes the linear order in the displacements. To
this order the (linear) strain field, $\mbox{\boldmath $\epsilon$}^{\mathrm{{lin}}}$ is given by:
\begin{equation}
\epsilon^{\mathrm{{lin}}}_{\alpha\beta}({\bf r},t)=\frac{1}{2}\bigg(\frac{\partial 
 u_\alpha ^{\mathrm{{lin}}}({\bf r},t)}{\partial r_\beta}
+\frac{\partial 
 u_\beta ^{\mathrm{{lin}}}({\bf r},t)}{\partial r_\alpha}\bigg)
\label{epslin}
\end{equation}
Keeping in Eq.~(\ref{slow1}) only terms that are quadratic in the 
displacements  one obtains (note that the potential is of second
order in the displacements): 
\begin{equation}
 \frac{\partial {e}^{\mathrm{{pot}}}} {\partial {t}}
= -\mathrm{div}\,{{\bf Q}}^{\mathrm{{force}}}
 + \frac{\partial {\bf {V}}_\beta}
{\partial r_\alpha}
  {\sigma}^{\mathrm{ {cont}}}_{\beta\alpha}
-\rho {\Gamma}\label{slow2.5}
\end{equation}
Since for quadratic potentials $\mbox{\boldmath $\sigma$}^{\mathrm{cont}}$ is
symmetric (as is easy to check, see also Eq.~(\ref{eq:lin_stress})) one can rewrite
Eq.~(\ref{slow2.5}) as follows:
\begin{equation}
 \frac{\partial {e}^{\mathrm{{pot}}}} {\partial {t}}
= -\mathrm{div}\,{{\bf Q}}^{\mathrm{{force}}}
 + \frac{1}{2} \bigg( \frac{\partial {\bf {V}}_\beta}
{\partial r_\alpha}
+ \frac{\partial {\bf {V}}_\alpha}
{\partial r_\beta}\bigg)
  {\sigma}^{\mathrm{ {cont}}}_{\beta\alpha}
- \rho {\Gamma}\label{slow2}
\end{equation}
hence (to quadratic  order in the displacements)
\begin{equation}
 \frac{\partial {e}^{\mathrm{{pot}}}} {\partial {t}}
= -\mathrm{div}\, {{\bf Q}}^{\mathrm{{force}}}
 + 
\frac{\partial \epsilon^{\mathrm{{lin}}}_{\alpha\beta}} {\partial {t}}
  {\sigma}^{\mathrm{ {cont}}}_{\beta\alpha}
- \rho {\Gamma}\label{slow3}
\end{equation}
Consider next the following virtual dynamical evolution from a force free
initial state to a final state at time $t_{\mathrm{{f}}}$:
 ${\bf u}_i(t) = {\bf u}_i(t_{\mathrm{{f}}}) 
\frac{t}{t_{\mathrm{{f}}}}$.
Assume that the reversible part of the force,
 ${\bf f}^{\mathrm{{rev}}}_{i/j}$ is linear in the displacement difference ${\bf u}_{ij}$ and
the irreversible part of the force is linear in the velocity difference
${\bf v}_{ij}$.  In this
dynamics the velocities are constant, the strain field satisfies
$\mbox{\boldmath$\epsilon$}^{\mathrm{{lin}}}({\bf r},t)=
\mbox{\boldmath$\epsilon$}^{\mathrm{{lin}}}({\bf r},t_{\mathrm{{f}}})\frac{t}{t_{\mathrm{{f}}}}$, the reversible part of the contact stress,
$\mbox{\boldmath$\sigma$}^{\mathrm{{cont,rev}}}$ (in which the force
is replaced by the reversible part of the force) satisfies
$\mbox{\boldmath$\sigma$}^{\mathrm{{cont,rev}}}({\bf r},t) =
\mbox{\boldmath$\sigma$}^{\mathrm{{cont,rev}}}({\bf r},t_{\mathrm{{f}}}) 
\frac{t}{t_{\mathrm{{f}}}}$, the irreversible part of the contact stress,
$\mbox{\boldmath$\sigma$}^{\mathrm{{cont,irrev}}}({\bf r},t)$, is constant in time, and similarly the  part of the heat flux which corresponds to the reversible parts of the
forces satisfies
${\bf Q}^{\mathrm{force,rev}}({\bf r},t)=
{\bf Q}^{\mathrm{force,rev}}({\bf r},t_{\mathrm{{f}}})
\frac{t}{t_{\mathrm{{f}}}}$ and 
${\bf Q}^{\mathrm{force,irrev}}({\bf r},t)$ is constant in time. In addition,
$\rho \Gamma({\bf r},t)$ is constant in time. Note that within the 
approximation in which only the lowest nonvanishing  order in the 
displacement is retained, 
${\bf r}_{ij}(t)$ is replaced by   ${\bf r}_{ij}(0)$. Upon substituting
these expressions in Eq.~({\ref{slow3}) and integrating Eq.~(\ref{slow3})
over time from zero to $t_{\mathrm{{f}}}$ one obtains: 
\begin{displaymath}
e({\bf r}, t_{\mathrm{{f}}}) = \frac{1}{2} \epsilon^{\mathrm{{lin}}}_{\alpha\beta}
({\bf r}, t_{\mathrm{{f}}}) \sigma^{\mathrm{{cont,rev}}}_{\beta\alpha}
({\bf r}, t_{\mathrm{{f}}})
+ \epsilon^{\mathrm{{lin}}}_{\alpha\beta}
({\bf r}, t_{\mathrm{{f}}}) \sigma^{\mathrm{{cont,irrev}}}_{\beta\alpha}
({\bf r}, t_{\mathrm{{f}}}) 
\end{displaymath}
\begin{equation}
-\frac{1}{2}\mathrm{div}\, {\bf Q}^{\mathrm{{force,rev}}}({\bf r}, t_{\mathrm{{f}}}) t_{\mathrm{{f}}}
-\mathrm{div}\,{\bf Q}^{\mathrm{{force,irrev}}}({\bf r}, t_{\mathrm{{f}}})
{ t_{\mathrm{{f}}}}
-\rho({\bf r}, t_{\mathrm{{f}}}) \Gamma({\bf r}, t_{\mathrm{{f}}})
{ t_{\mathrm{{f}}}} \label{almost}
\end{equation}
In order  to keep the displacements fixed we replace the velocities
that appear in the expressions for the various fields by $1/t_{\mathrm{{f}}}$ times the corresponding displacements. A field in which the velocities have
been replaced by the corresponding  displacements is marked by a superscript I. It follows that at $t=t_{\mathrm{{f}}}$:
$\mbox{\boldmath$\sigma$}^{\mathrm{{cont,rev,I}}}=\mbox{\boldmath$\sigma$}^{\mathrm{{cont,rev}}}$,
$\mbox{\boldmath$\sigma$}^{\mathrm{{cont,irrev,I}}} = t_{\mathrm{{f}}}
\mbox{\boldmath$\sigma$}^{\mathrm{{cont,rev}}}$, 
${\bf Q}^{\mathrm{{force,rev,I}}}= t_{\mathrm{{f}}}
{\bf Q}^{\mathrm{{force,rev}}}$,
${\bf Q}^{\mathrm{{force,irrev,I}}}= t^2_{\mathrm{{f}}}{\bf Q}^{\mathrm{{force,irrev}}}$ (recall the assumption that ${\bf f}^{\mathrm{{irrev}}}_{i/j}$
 is linear
 in the particle velocities), 
and $(\rho\Gamma)^{\mathrm{{I}}} = t^2_{\mathrm{{f}}} \rho\Gamma$. 
Substituting these expressions in Eq.~(\ref{almost}), keeping the displacements
fixed and letting $t_{\mathrm{{f}}}$ go to infinity, one obtains: 
\begin{equation}
e({\bf r}, t_{\mathrm{{f}}}) = \frac{1}{2} \epsilon^{\mathrm{{lin}}}_{\alpha\beta}
({\bf r}, t_{\mathrm{{f}}}) \sigma^{\mathrm{{cont,rev}}}_{\beta\alpha}
({\bf r}, t_{\mathrm{{f}}})
-\frac{1}{2}\mathrm{div}\, {\bf Q}^{\mathrm{{force,rev,I}}}({\bf r}, t_{\mathrm{{f}}}) 
 \label{tiny}
\end{equation}
where (changing $t_{\mathrm{{f}}}$ to $t$) if follows from the expression
for the heat flux, Eq~(\ref{heatflux1}) and the above definitions that
 \begin{eqnarray}
Q^{\mathrm{{force,rev,I}}}_\alpha &=&
\frac{1}{4} \sum_{ij} {\bf f}^{\mathrm{{rev}}}_{ij}(t)\cdot \left( {\bf u}^\prime_i({\bf r},t)
+ {\bf u}^\prime_j({\bf r},t) \right) r_{ij\alpha}(0)\\
&&\times  \int_0^1 ds \phi\left[{\bf r}-{\bf r}_i(t) + s {\bf r}_{ij}(t)\right]
\label{last}
\end{eqnarray}
where 
\begin{equation}
\label{flu}
{\bf u}^\prime_i({\bf r},t) \equiv {\bf u}_i(t) - {\bf u}^{\mathrm{{lin}}}
({\bf r},t)
\end{equation}
is the displacement fluctuation of particle $i$. 
We have thus obtained a correction to the classical formula for the
elastic energy which   represents the fluctuating  work and stems from the integral
of the heat flux over time. This result can be directly obtained from
an analysis of the elastic energy, cf.  Eq.~(\ref{eee}).

\end{document}